\documentclass[pre,twocolumn,showpacs]{revtex4}

\usepackage{graphicx}
\usepackage{amsmath}
\usepackage{mathtools}
\usepackage{amssymb}

\usepackage{verbatim}   % useful for program listings
\usepackage{color}      % use if color is used in text
\usepackage{subfigure}  % use for side-by-side figures
\usepackage{hyperref}   % use for hypertext links, including those to external documents and URLs

\usepackage{soul,xcolor}

\newcommand{\be}{\begin{equation}}
\newcommand{\ee}{\end{equation}}
\newcommand{\ba}{\begin{eqnarray}}
\newcommand{\ea}{\end{eqnarray}}

\newcommand{\ra}{\rho({\bf r})}
\newcommand{\rb}{\rho({\bf r}')}
\newcommand{\rc}{\rho({\bf r}'')}
\newcommand{\ua}{u({\bf r},{\bf r}')}

\newcommand{\inta}{\int d{\bf r}}
\newcommand{\intb}{\int d{\bf r}'}
\newcommand{\intc}{\int d{\bf r}''}

\begin{document}

%==============================================================
\title{Inhomogeneous fluid of penetrable-spheres:  application of the random phase approximation}
%==============================================================

\author{Yan Xiang}
\email{hnxxxy123@sjtu.edu.cn}
%\homepage{}
\affiliation{School of Chemistry and Chemical Engineering, Shanghai Jiao Tong University, 
Shanghai 200240, China}

\author{Derek Frydel}
\email{dfrydel@gmail.com}
%\homepage{}
\affiliation{Institute of Physics, The Federal University of Rio Grande do Sul, Porto Alegre 91501-970, Brazil}

\date{\today}

\begin{abstract}
The focus of the present work is the application of the random phase approximation (RPA), derived for inhomogeneous 
fluids [Frydel and Ma, Phys. Rev. E 93, 062112 (2016)], to penetrable-spheres. As penetrable-spheres transform into 
hard-spheres with increasing interactions, they provide an interesting case for exploring the RPA, its shortcomings, and 
limitations, the weak- versus the strong-coupling limit. Two scenarios taken up by the present study are a one-component 
and a two-component fluid with symmetric interactions. In the latter case, the mean-field contributions cancel out and any 
contributions from particle interactions are accounted for by correlations. The accuracy of the RPA for this case is the result 
of a somewhat lucky cancellation of errors.
%The focus of the present work is the application of the random phase approximation (RPA), derived in Ref. [D. Frydel and 
%M. Ma, Phys. Rev. E {\bf 93}, 062112 (2016)] for inhomogeneous fluids, to penetrable-spheres.  As penetrable-spheres 
%transform into hard-spheres with increasing interactions, they provide an interesting case for exploring the RPA, 
%its shortcomings and limitations, the weak- versus the strong-coupling limit.  Two scenarios taken up by the present study 
%are a one-component and a two-component fluids with symmetric interactions.  It is the latter case for which the RPA is 
%particularly useful.  In this case the mean-field contributions are cancelled and any contributions from particle interactions 
%are accounted for by correlations.  The accuracy of the RPA for this case is the result of a somewhat lucky cancellation of 
%errors.  
\end{abstract}

\pacs{
}

\maketitle
%------------------------------------------------

\section{Introduction}
Recently in Ref. 1 we have extended the random phase approximation (RPA) to inhomogeneous fluids as a systematic 
attempt to supplement the mean-field description with correlations \cite{Frydel16b}. When that approximation is applied 
to Coulomb particles, one finds the same set of equations as those constituting the variational Gaussian approximation 
within the field-theoretical framework generated by transforming the original partition function into a functional integral
\cite{Netz00,Wang10,Tony14,David14,Frydel15}.
(The variational Gaussian approximation for ions consists of two coupled equations. One is the Poisson-Boltzmann equation 
with a modified charge density that includes correlational contributions. And the second equation determines correlational 
contributions and can be represented as the Ornstein-Zernike (OZ) equation, suggesting an alternative derivation within 
the density functional theory.) It then follows that the variational Gaussian approximation is nothing other than the RPA 
approximation formulated within a different framework. The density functional formulation, however, not only offers an 
alternative derivation. Its greatest merit is that it provides generalization to any particle system as the formulation makes 
no assumptions about pair interactions. The question then is not for what pair interactions, but rather under what conditions 
the RPA is reliable. The conditions at which the RPA fails for penetrable-spheres should be general and apply to Coulomb 
particles or the Gaussian core model. Defining the limits of the validity of the RPA for one system, therefore, offers insights 
about those limits in another system.

Because the RPA approach of Ref. \cite{Frydel16} is non-perturbative (correlations and density are obtained self-consistently), it 
is tempting to think it is suitable for describing the strong-coupling limit. However, the study of Ref. \cite{Frydel16} for a number of 
pair interactions does not support this conjecture. If anything, it suggests the contrary view that the RPA, despite its non-perturbative 
feature, is strictly a theory of weakly correlated systems. Again, by identifying the variational Gaussian approximation of the 
field-theoretical formulation as the RPA, this conclusion is not surprising. The RPA has been used for long time in the electronic 
structure DFT where it?s known as a theory of the weakly-correlated limit \cite{Ren12}. One goal of this paper, therefore, is to dispel 
the misunderstanding that the variational Gaussian approximation has potential for describing the strong-coupling regime of a 
charged or any system.
%Because the RPA approach of Ref. \cite{Frydel16} is non-perturbative (correlations and density are obtained self-consistently), 
%it is tempting to think it is suitable for describing the strong-coupling limit.  However, the study of Ref. \cite{Frydel16} for different 
%pair interactions does not support this conjecture.  If anything, it suggests the contrary view, that the RPA, despite its 
%non-perturbative feature, is strictly a theory of weakly correlated systems.  Again, by identifying the variational Gaussian 
%approximation of the field-theoretical formulation as the RPA, this conjectures is not at all surprising.  The RPA has a lengthy history 
%of being used in the electronic structure DFT.  Researchers in that area know all along that the RPA is not the theory of the 
%strong-coupling limit \cite{Ren12}.  One goal of this paper, therefore, it to dispel the misunderstanding that the variational 
%Gaussian approximation has potential for describing the strong-coupling regime of a charged or any system.  

In Ref. \cite{Frydel16} we applied the RPA to Coulomb particles and the Gaussian core model (considered to be a weakly correlated 
fluid \cite{Hansen00a,Evans01,Frydel16c}). In the present work, we continue to explore the RPA by its application to penetrable-spheres. 
Unlike hard-spheres, penetrable-spheres may overlap, however, every overlap incurs energy cost. As overlaps become more costly, 
penetrable-spheres transform into hard-spheres. As the RPA is not expected to work for hard-spheres, one should observe a gradual 
breakdown of the RPA as a function of increasing interactions. The case of penetrable-spheres, therefore, provides a useful testing 
ground for the performance of the RPA.
%In Ref. \cite{Frydel16} we applied the RPA to Coulomb particles and the Gaussian core model (an example of penetrable particles
%constituting a weakly correlated fluid \cite{Hansen00a,Evans01,Frydel16c}).  In the present work we continue to explore the RPA by 
%its application to penetrable-spheres.  Unlike hard-spheres, penetrable-spheres may overlap, however, every overlap incurs energy 
%cost.  As overlaps become more costly, penetrable-spheres transform into hard-spheres.  As the RPA is not expected to work for 
%hard-spheres, one should observe a gradual breakdown of the RPA as a function of an increasing overlap energy.  The case of 
%penetrable-spheres, therefore, provides an a useful testing ground for the performance of RPA.  

In Sec. II we lay down the framework of the RPA approximation. In Sec. \ref{sec:hom1} we investigate a homogenous case of 
penetrable-spheres. In Sec. \ref{sec:inhom1} we show the results for an inhomogeneous fluid near a planar wall. In Sec. \ref{sec:alter} 
we consider alternative (dilute limit) approaches and compare them with the RPA. In Sec. \ref{sec:hom2} we study a two-component 
penetrable-sphere fluid with symmetric interactions. Finally, in Sec. \ref{sec:concl} we conclude the work.
%In Sec. \ref{sec:RPA} we lay down the framework of the RPA approximation.  In Sec. \ref{sec:hom1} we investigate a homogenous 
%case of penetrable-spheres.  In Sec. \ref{sec:inhom1} we show the results for an inhomogeneous penetrable-spheres fluid near a 
%planar wall.  In Sec. \ref{sec:alter} we consider alternative (dilute limit) approaches and compare them with the RPA.  In Sec. 
%\ref{sec:hom2} we study a two-component penetrable-sphere fluid with symmetric interactions.  Finally in Sec. \ref{sec:concl} we 
%conclude the work.  

\section{The random phase approximation}
\label{sec:RPA}

In this section we go over the construction of the RPA approximation. 
We begin with the expression for the mean-field free energy,  
\ba
F_{\rm mf}[\rho] &=& F_{\rm id}[\rho] + \int \!d{\bf r}\,\rho({\bf r})U({\bf r}) \nonumber\\
&+& \frac{1}{2}\int \!d{\bf r}\!\int \!d{\bf r}'\ra\rb\ua.%\nonumber\\
\label{eq:Fmf}
\ea
The last term captures particle interactions, however, it neglects correlations which complete the free energy expression as
\be
F[\rho] = F_{\rm mf}[\rho] + F_c[\rho].  
\ee
The exact expression for $F_c[\rho]$ is obtained using the adiabatic connection wherein pair interactions $\lambda\ua$ are 
gradually switched on as $\lambda$ varies from $0\to 1$.  During the procedure the density $\rho({\bf r})$ is kept fixed by means 
of an external $\lambda$-dependent potential, $U_{\lambda}({\bf r})$, which in the limit $\lambda\to1$ recovers a true external 
potential.  Keeping a density independent of $\lambda$ is essential for constructing a density functional formulation.     

Defining the free energy for a given $\lambda$ as $F_{\lambda}=-k_BT\log Z_{\lambda}$ where 
\be
Z_{\lambda} = \frac{1}{N!\Lambda^{3N}}\!\!\int \!\!d{\bf r}_1\!\dots\!\!\int\!\!d{\bf r}_N\,
e^{-\beta \sum_iU_{\lambda}({\bf r}_i)}e^{-\beta \lambda\sum_{j>i}u({\bf r}_i,{\bf r}_j)}, 
\ee
and $N$ is the total number of particles, the free energy of a physical system at $\lambda=1$ is given by 
\be
F = F_{\lambda=0} + \int_0^1d\lambda\,\frac{\partial F_{\lambda}}{\partial\lambda}.  
\ee
After some substitution and manipulation we get the free energy expression where the correlational part is found to be 
\be
F_c[\rho] = \frac{1}{2}\int_0^1 d\lambda\inta\,\ra\intb\,\rb h_{\lambda}({\bf r},{\bf r}')\ua, 
\label{eq:Fc}
\ee
where $h_{\lambda}({\bf r},{\bf r}')$ is the correlation function of a fluid \cite{Hansen} with interactions $\lambda\ua$ and 
in the presence of the external potential $U_{\lambda}({\bf r})$, which although absent in Eq. (\ref{eq:Fc}) is implied by 
a fixed $\rho({\bf r})$.  

To complete the framework we still need a way to calculate $h_{\lambda}({\bf r},{\bf r}')$. To this end we introduce the 
Ornstein-Zernike equation (OZ), 
\be
h_{\lambda}({\bf r},{\bf r}') = 
c_{\lambda}({\bf r},{\bf r}') + \intc\,\rc h_{\lambda}({\bf r}',{\bf r}'')c_{\lambda}({\bf r},{\bf r}''),
\label{eq:OZ_lambda}
\ee
which expresses $h_{\lambda}({\bf r},{\bf r}')$ in terms of the direct correlation function $c_{\lambda}({\bf r},{\bf r}')$.  
A direct correlation function is obtained from the exact definition, 
\be
c_{\lambda}({\bf r},{\bf r}') \equiv -\beta \frac{\delta^2 F_{ex}^{\lambda}[\rho]}{\delta\rho({\bf r})\delta\rho({\bf r}')}, 
\label{eq:c_exact}
\ee
and since we are interested in a systematic method beyond the mean-field, we use the mean-field result for 
the excess free energy, $F_{\rm ex}\approx \frac{1}{2}\inta \intb \ra\rb\ua$, which yields 
\be
c_{\lambda}^{\rm rpa}({\bf r},{\bf r}')=-\lambda \beta\ua.  
\ee
The above is known as the random phase approximation closure.  Inserting this into the OZ equation we get 
\ba
h_{\lambda}^{\rm rpa}({\bf r},{\bf r}') &=& -\lambda\beta u({\bf r},{\bf r}') \nonumber\\
&-&\lambda\beta \int d{\bf r''}\,\rho({\bf r}'')h_{\lambda}^{\rm rpa}({\bf r}'',{\bf r}')u({\bf r},{\bf r}''),
\label{eq:OZ_RPA}
\ea
and the correlation free energy becomes
\be
F_{c}^{\rm rpa}[\rho] = - \frac{1}{2}\inta\ra\int_0^1 d\lambda\,\frac{h_{\lambda}^{\rm rpa}({\bf r},{\bf r})+\lambda\beta u({\bf r},{\bf r})}{\lambda\beta}.  
\label{eq:Fc_rpa}
\ee
$h_{\lambda}^{\rm rpa}({\bf r},{\bf r}')$ can alternatively be represented as an expansion 
\ba
&&h_{\lambda}^{\rm rpa}({\bf r},{\bf r}') =-\beta\lambda u({\bf r},{\bf r}') \nonumber\\
&+&\beta^2\lambda^2\int \!\!d{\bf r}_1\,\rho({\bf r}_1) u({\bf r},{\bf r}_1)u({\bf r}_1,{\bf r}')\nonumber\\
&-& \beta^3\lambda^3\int \!\!d{\bf r}_1\!\!\int \!\!d{\bf r}_2\,
\rho({\bf r}_1)\rho({\bf r}_2)u({\bf r},{\bf r}_1)u({\bf r}_1,{\bf r}_2)u({\bf r}_2,{\bf r}')\nonumber\\ 
&+& \dots
\label{eq:hL}
\ea
generated by repetitive substitution of $h_{\lambda}^{\rm }({\bf r},{\bf r}')$ each time it appears on the right-hand side
of the OZ equation.  

Using the above expansion, the correlation free energy becomes 
\ba
\beta F_c^{\rm rpa}[\rho] &=& -\frac{\beta^2}{4}\!\!\int \!\!d1\!\!\int \!\!d2\,\rho_1\rho_2u_{12}u_{21}\nonumber\\
&+&\frac{\beta^3}{6}\!\!\int \!\!d1\!\!\int \!\!d2\!\!\int \!\!d3\, \rho_1\rho_2\rho_3 u_{12}u_{23}u_{31}\nonumber\\
&-&\frac{\beta^4}{8}\!\!\int \!\!d1\!\!\int \!\!d2\!\!\int \!\!d3\!\!\int \!\!d4\, \rho_1\rho_2\rho_3 \rho_4 u_{12}u_{23}u_{34}u_{41}\nonumber\\
&+&\dots
\ea
where $\rho({\bf r}_i)\to\rho_i$, $u({\bf r}_i,{\bf r}_j)\to u_{ij}$, and $\int d{\bf r}_i\to \int di$. 
All the constituent integral terms have a ring topology, a mathematical signature of the RPA.  

The above expansion is handy for generating functional derivatives.  For example, a true density satisfies the relation 
$\frac{\delta F}{\delta\rho}=\mu$, where $\mu$ denotes the chemical potential.  A functional derivative of $F_c$ with respect 
to $\rho({\bf r})$ is carried out term by term, leading to a new expansion that is identified as  
\be
\frac{\delta \beta F_c^{\rm rpa}[\rho]}{\delta\rho({\bf r})} = -\frac{1}{2}\bigg(h^{\rm rpa}({\bf r},{\bf r}) + \beta u({\bf r},{\bf r})\bigg).  
\ee
A physical density then satisfies 
\ba
\rho^{\rm }({\bf r}) &=& \rho_b e^{-\beta U({\bf r})}e^{-\beta\intb\,[\rho^{\rm }({\bf r}')-\rho_b]\ua}\nonumber\\
&\times&e^{\frac{1}{2}[h^{\rm rpa}({\bf r},{\bf r})-h^{\rm rpa}_b(0)]}, 
\label{eq:rho_rpa}
\ea
where $\rho_b$ is the bulk density, and $h_b(0)=h_b({\bf r},{\bf r})$ is the correlation function in a bulk.   
The correlation function is obtained from the OZ equation in Eq. (\ref{eq:OZ_RPA}).  Because the two equations are coupled
the method is self-consistent (as opposed to perturbative).  

In the introduction it was said that the RPA yields the same set of equations as the  
variational Gaussian approximation for Coulomb charges \cite{Netz00,Wang10}.   Since Eq. (\ref{eq:OZ_RPA}) and (\ref{eq:rho_rpa}) 
are the two coupled equations of the RPA approximation, by substituting for $u(r)$ the Coulomb interactions $\sim 1/r$ 
we should recover the equations of the variational Gaussian approximation.  Indeed, this turns out to be the case.

Borrowing from the conventions of the linear algebra, the RPA formulas can be expressed more compactly.
By introducing the operator $A({\bf r},{\bf r}')=\rho({\bf r})\beta \ua$ and multiplication 
$A^2({\bf r},{\bf r}') = \int d{\bf r}''\,A({\bf r},{\bf r}'')A({\bf r}'',{\bf r}')$ and so forth, the correlation function becomes
\be
\rho({\bf r})h_{\lambda}^{\rm rpa}({\bf r},{\bf r}') = -\frac{\lambda A}{I+\lambda A},
\ee
where $I=\delta({\bf r}-{\bf r}')$ represents the identity matrix in the continuous limit, and the correlation free energy is 
\be
\beta F_c^{\rm rpa} = \frac{1}{2}{\rm Tr}\,\bigg(\log[I+A] - A\bigg).  
\ee
In the language of integrals the trace of an operator $A({\bf r},{\bf r}')$ is defined as 
$$
{\rm Tr}\, A = \int d{\bf r}\, A({\bf r},{\bf r}).  
$$

In this article we apply the RPA method laid down above to penetrable-spheres, whose interactions, 
\be
\beta u({\bf r},{\bf r}') = \varepsilon\theta(\sigma - |{\bf r}-{\bf r}'|),
\ee
are given in terms of the Heaviside function $\theta(x)$, such that $\theta(x>0)=1$ and $\theta(x<0)=0$, 
and $\varepsilon$ is the interaction strength.  

A quick glance at Fig. (\ref{fig:hr}) reveals discontinuity in the pair correlation function.  From the expansion in 
Eq. (\ref{eq:hL}) we know that the discontinuity in $h_{\lambda}^{\rm rpa}({\bf r},{\bf r}')$ comes from the first term, that is, 
from interactions.  The second term is already continuous and represents an overlap between two spheres of radius $\sigma$.  
Within the RPA the quantity $h^{\rm rpa}({\bf r},{\bf r}')+\varepsilon\theta(\sigma - |{\bf r}-{\bf r}'|)$, therefore, is continuous.

\section{A Homogenous scenario}
\label{sec:hom1}

In the present section we consider a homogenous fluid.  As an example of a quantity that depends on correlations 
we consider an average potential energy per particle, 
\be
\beta E  = 4\pi\varepsilon\rho_b\int_0^{\sigma} dr\,r^2 \bigg[1+h_b(r)\bigg].  
\label{eq:E}
\ee
For the RPA the above expression simplifies by using the OZ relation of Eq. (\ref{eq:OZ_RPA}) into
\be
\beta E = \frac{\alpha}{3} - h_b^{\rm rpa}(0)-\varepsilon, 
\label{eq:E_rpa}
\ee
where we introduce the dimensionless parameter $\alpha=4\pi\varepsilon\rho_b\sigma^3$.  The quantity 
$h^{\rm rpa}(0)$ is obtained by Fourier transforming the OZ equation in Eq. (\ref{eq:OZ_RPA}), yielding  
\be
\tilde h^{\rm rpa}(k) = -\frac{\beta \tilde u(k)}{1+\beta\rho_b \tilde u(k)}, 
\label{eq:hk}
\ee
where $\beta \tilde u(k) = 4\pi \varepsilon\sigma^3(\sin k\sigma - k\sigma\cos k\sigma)/(k\sigma)^3$ is the Fourier transformed 
pair potential of penetrable-spheres.  Performing the inverse transform we get
\be
h^{\rm rpa}(0) = -\bigg(\frac{1}{2\pi}\bigg)^{3}\int_0^{\infty} dk\, \frac{4\pi k^2 \beta \tilde u(k)}{1+\beta\rho_b \tilde u(k)},
\label{eq:h0}
\ee
and the integral is evaluated numerically.

\begin{figure}[h] 
 \begin{center}
 \begin{tabular}{rr}
  \includegraphics[height=0.2\textwidth,width=0.3\textwidth]{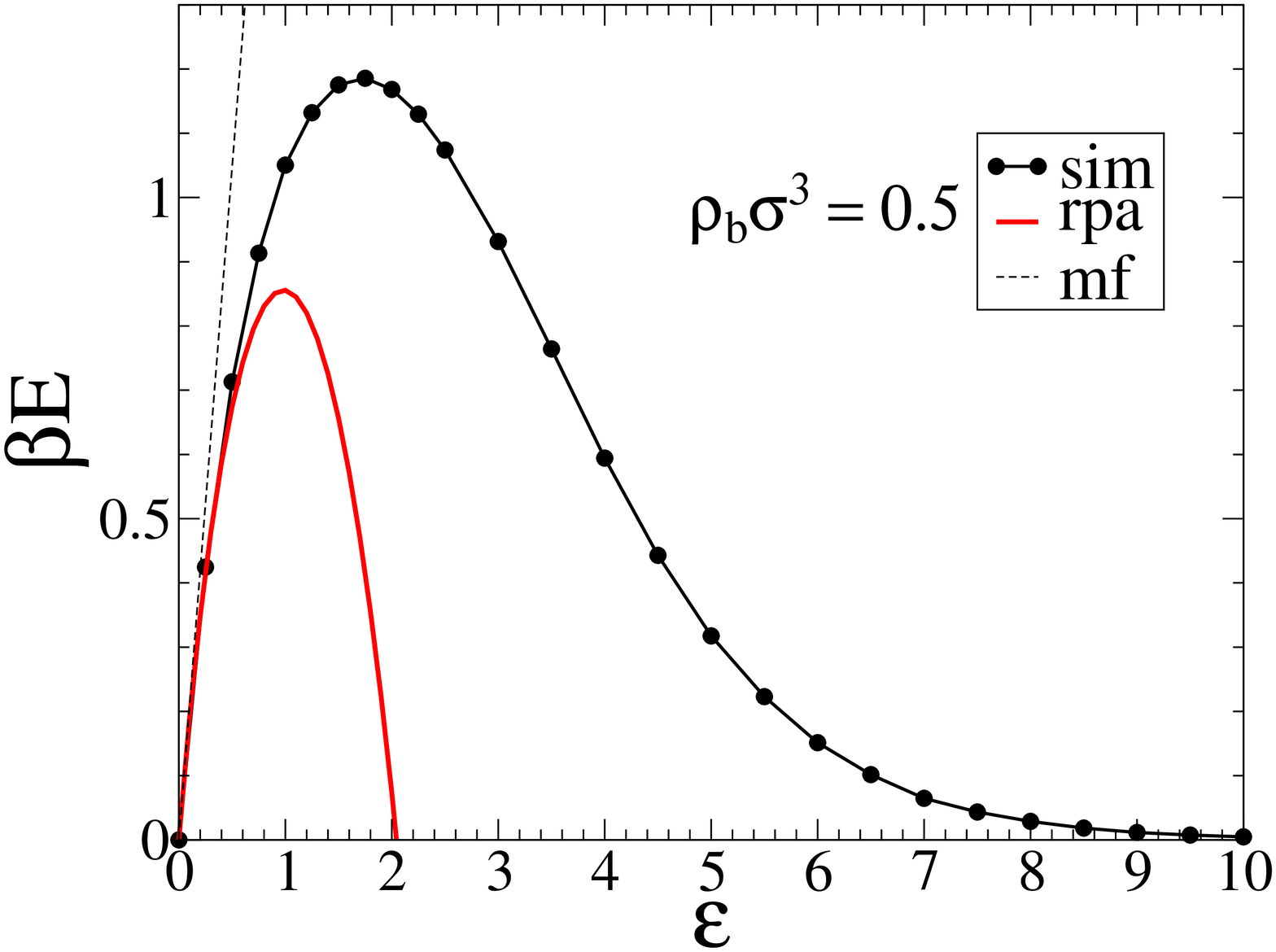}\\
 \end{tabular}
 \end{center}
\caption{Average potential energy per particle as a function of an interaction strength for a homogenous system for $\rho_b\sigma^3=0.5$.  
Exact results are from the MC simulation
for $N=1000$ particles.  The RPA data points correspond to Eq. (\ref{eq:E_rpa}), and the mean-field result to the first term of that expression.} 
\label{fig:E5}
\end{figure}

Fig. (\ref{fig:E5}) shows the data points for $\beta E$ (defined in Eq. (\ref{eq:E})) as a function of $\varepsilon$ for a fluid with 
density $\rho_b\sigma^3=0.5$.  The exact results show a non-monotonic behavior with a peak at $\varepsilon\approx 2$.  
The non-monotonic behavior may appear counterintuitive, since it is expected that $\beta E$ should increase with increasing repulsive 
interactions.  But in the case of penetrable-spheres large repulsive interactions also imply lower chance of overlaps between particles.  
If we turn next to the data points for the RPA, we note how its reliability is limited within the range $\varepsilon\le 0.5$, 
indicating poorly predicted correlations for higher values of $\varepsilon$.  
%%%%%%%%%%%%%%%%%%%%%
\graphicspath{{figures/}}
%%%%%%%%%%%%%%%%%%%%%

\begin{figure}[h] 
 \begin{center}
 \begin{tabular}{rr}
  \includegraphics[height=0.18\textwidth,width=0.22\textwidth]{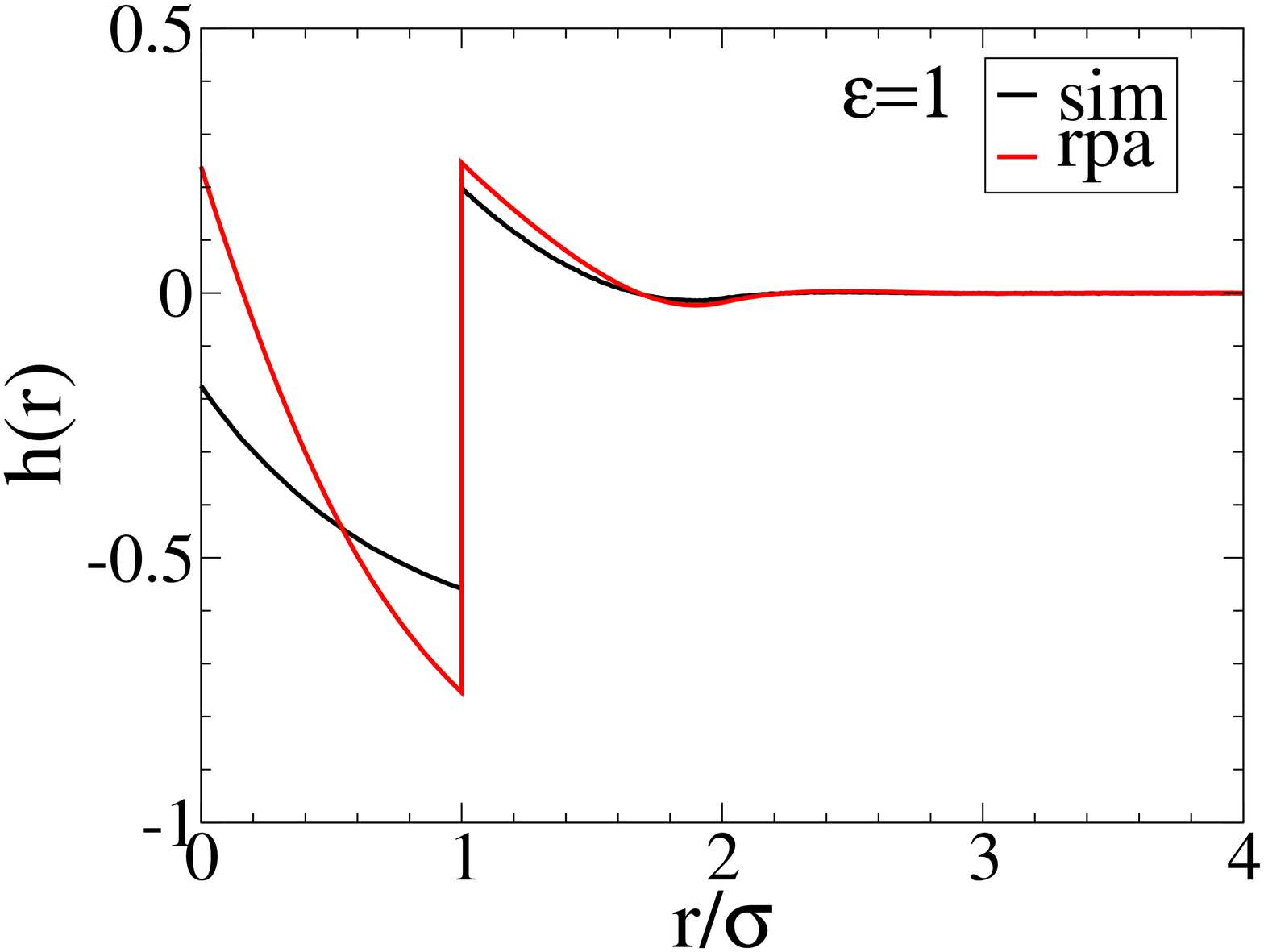}&
  \includegraphics[height=0.18\textwidth,width=0.22\textwidth]{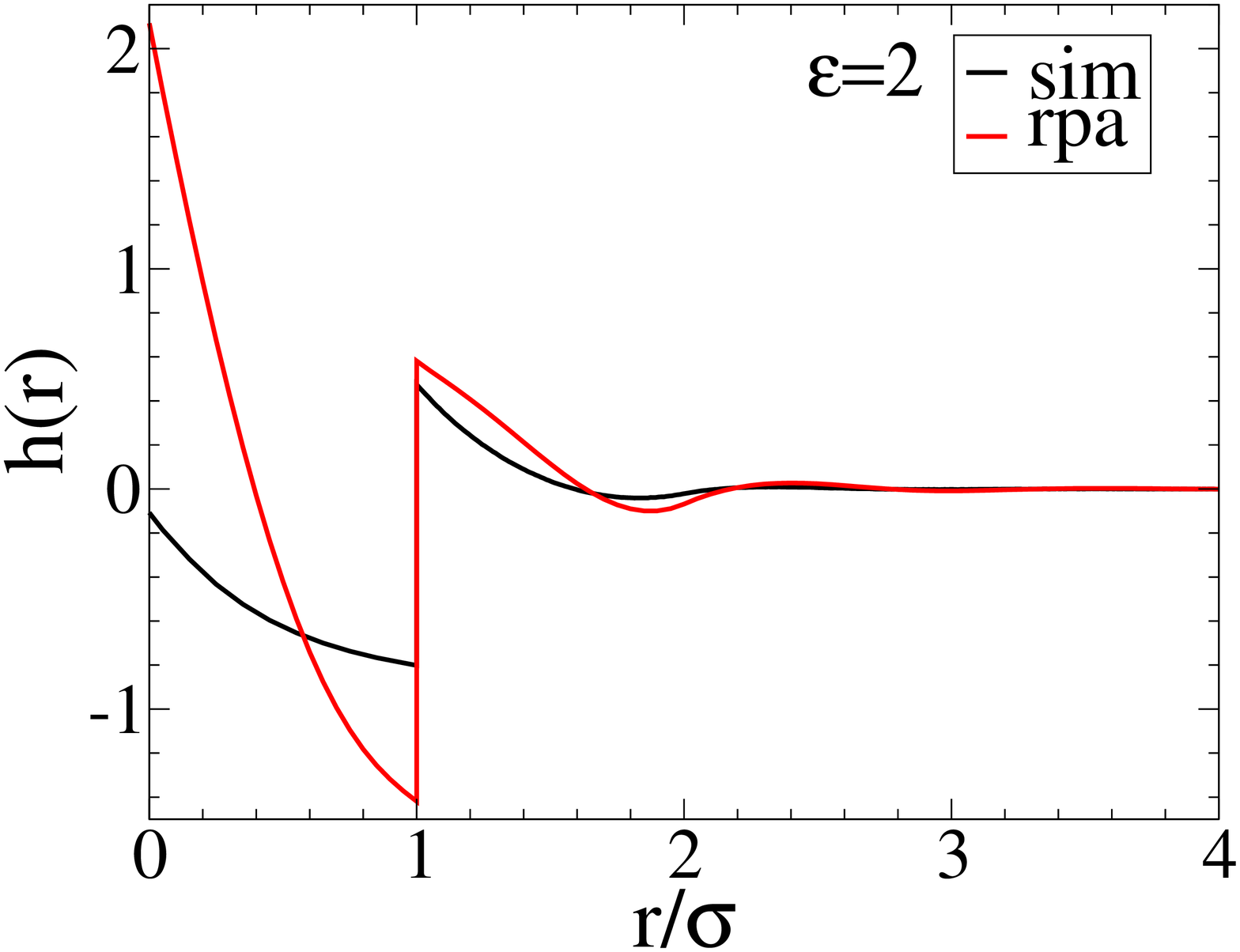}\\
 \end{tabular}
 \end{center}
\caption{Pair correlation functions of a homogeneous fluid for $\rho_b\sigma^3=0.5$.  } 
\label{fig:hr}
\end{figure}

To understand the above results, in Fig. (\ref{fig:hr}) we plot correlation functions which contribute to the behavior of $\beta E$.  
The RPA correlations are calculated numerically from 
\be
h(r) = -\bigg(\frac{1}{2\pi}\bigg)^{3}\int_0^{\infty} dk\, \frac{4\pi k^2 \beta \tilde u(k)}{1+\beta\rho_b \tilde u(k)}\frac{\sin kr}{kr}.  
\ee 
For separations $r>\sigma$ the resulting correlations compare reasonably well with the exact results.   
However, it is correlations at separations less than $\sigma$ that determine $\beta E$, and here the RPA correlations fail.  
%%%%%%%%%%%%%%%%%%%%%
\graphicspath{{figures/}}
%%%%%%%%%%%%%%%%%%%%%

\begin{figure}[h] 
 \begin{center}
 \begin{tabular}{rr}
  \includegraphics[height=0.18\textwidth,width=0.22\textwidth]{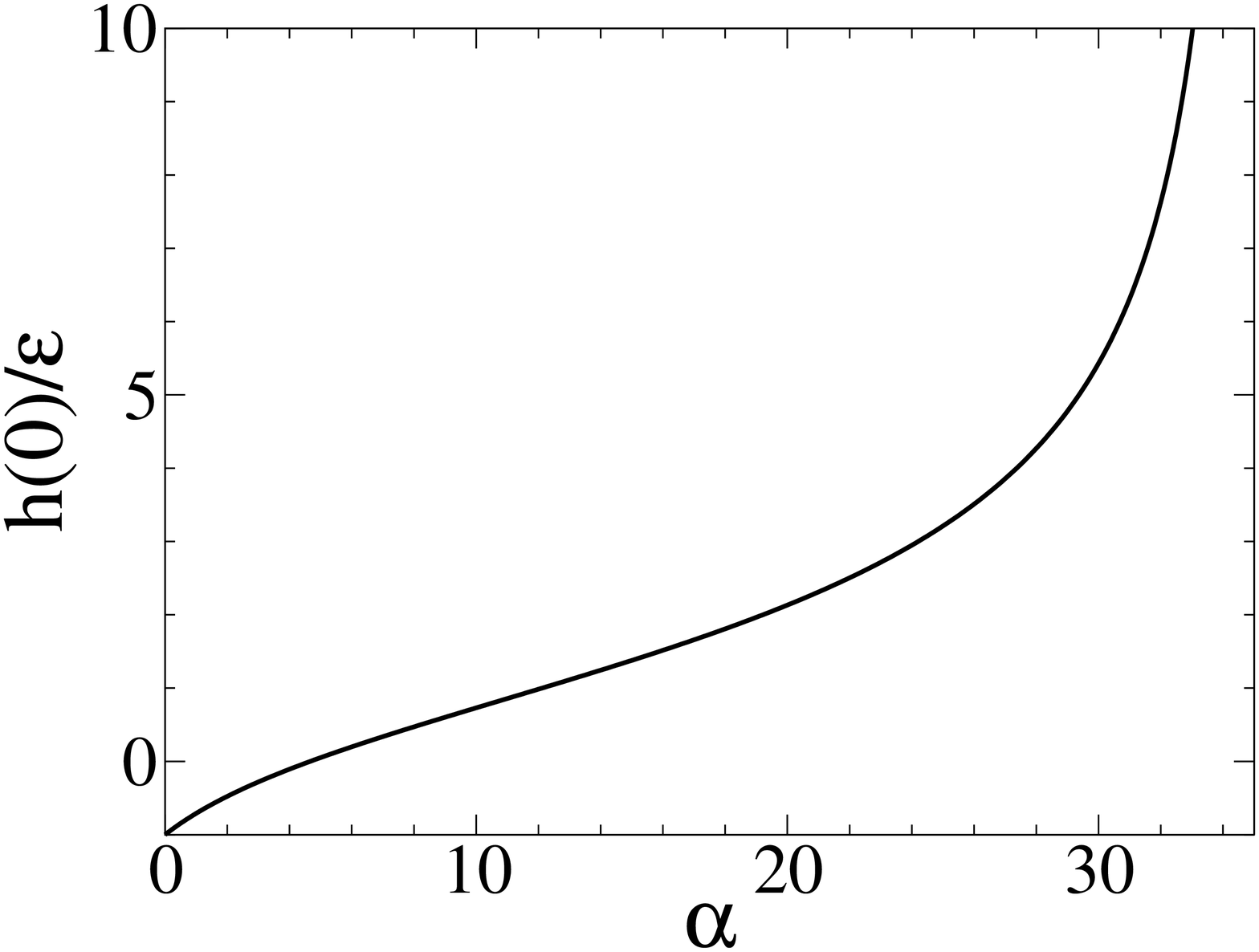}&
  \includegraphics[height=0.18\textwidth,width=0.22\textwidth]{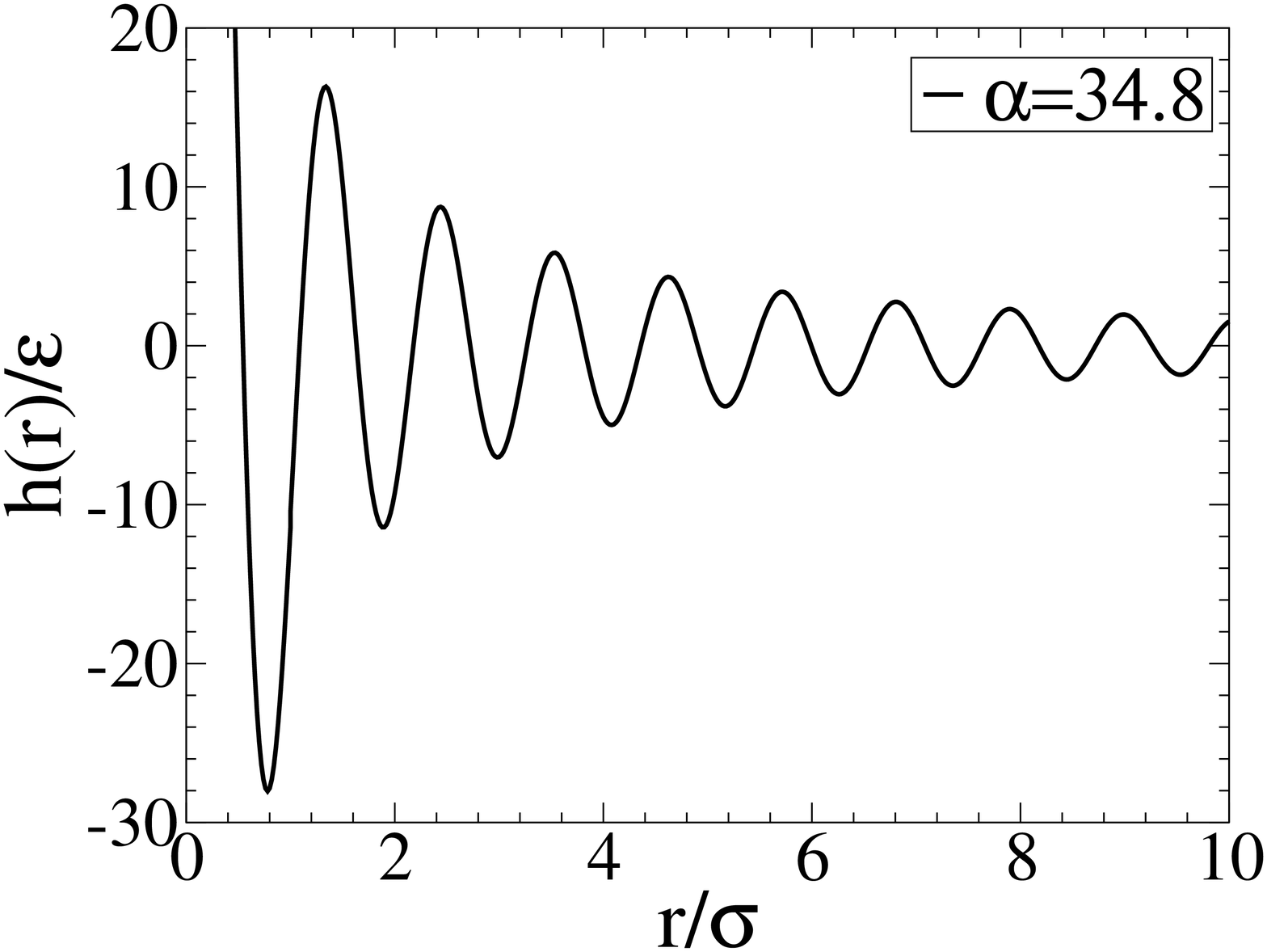}\\
 \end{tabular}
 \end{center}
\caption{(a) The correlation function for a separation r = 0 as a function of $\alpha=4\pi\varepsilon\rho_b\sigma^3$. 
(b) The correlation function for ?? near the point of divergence. }
\label{fig:h0}
\end{figure}

We next focus on a quantity $h^{\rm rpa}(0)$ as a function of $\alpha$ that determines $\beta E$ 
within the RPA (see Eq. (\ref{eq:E_rpa})).  The results are shown in Fig. (\ref{fig:h0}) (a) 
and indicate a sharp increase of $h^{\rm rpa}(0)$ and then as $\alpha\to 34.81$, $h^{\rm rpa}(0)$ diverges.  
%%%%%%%%%%%%%%%%%%%%%
\graphicspath{{figures/}}
%%%%%%%%%%%%%%%%%%%%%
Furthermore, the divergence is coextensive with the emergence of the solid like regular structure as shown in Fig. (b) and 
signaling the presence of an anomalous phase transition (anomalous because not predicted by simulations).  The divergence occurs  
when the denominator in the expression for $\tilde h^{\rm rpa}(k)$ vanishes, which can only happen for negative $\tilde u(k)$.  
(For the Gaussian core model where $\beta \tilde u^{\rm}(k)=\pi^{3/2}\varepsilon\sigma^3 e^{-k^2\sigma^2/4}\ge 0$ the divergence
never arises \cite{Lowen01}).

\begin{figure}[h!]
\centering
\subfigure[\ $\eta_b=0.1$]{
\includegraphics[scale=0.2]{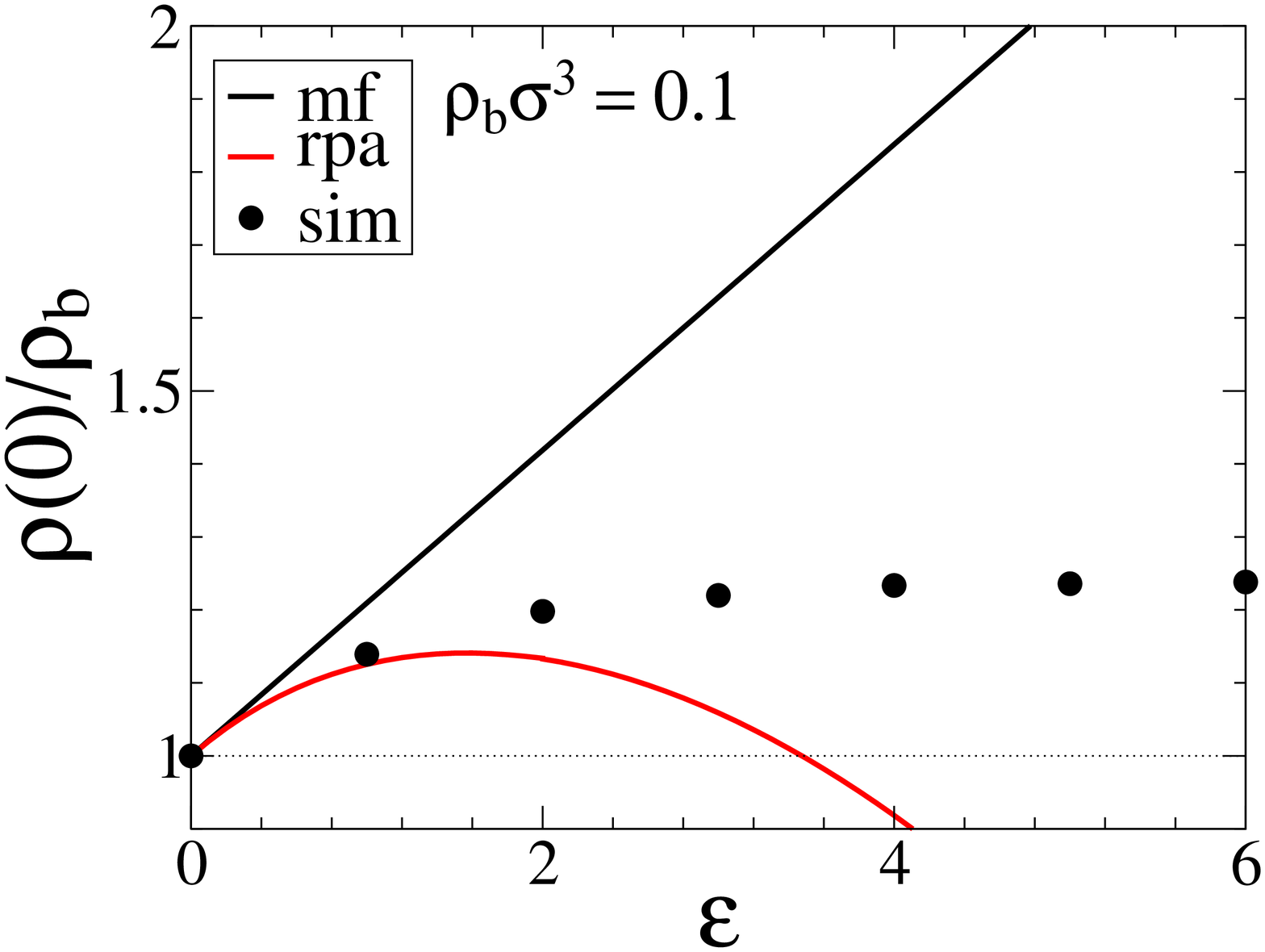}}
\subfigure[\ $\eta_b=0.3$]{
\includegraphics[scale=0.2]{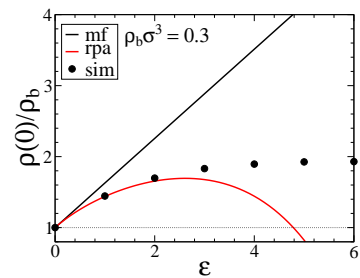}}
\subfigure[\ $\eta_b=0.5$]{
\includegraphics[scale=0.2]{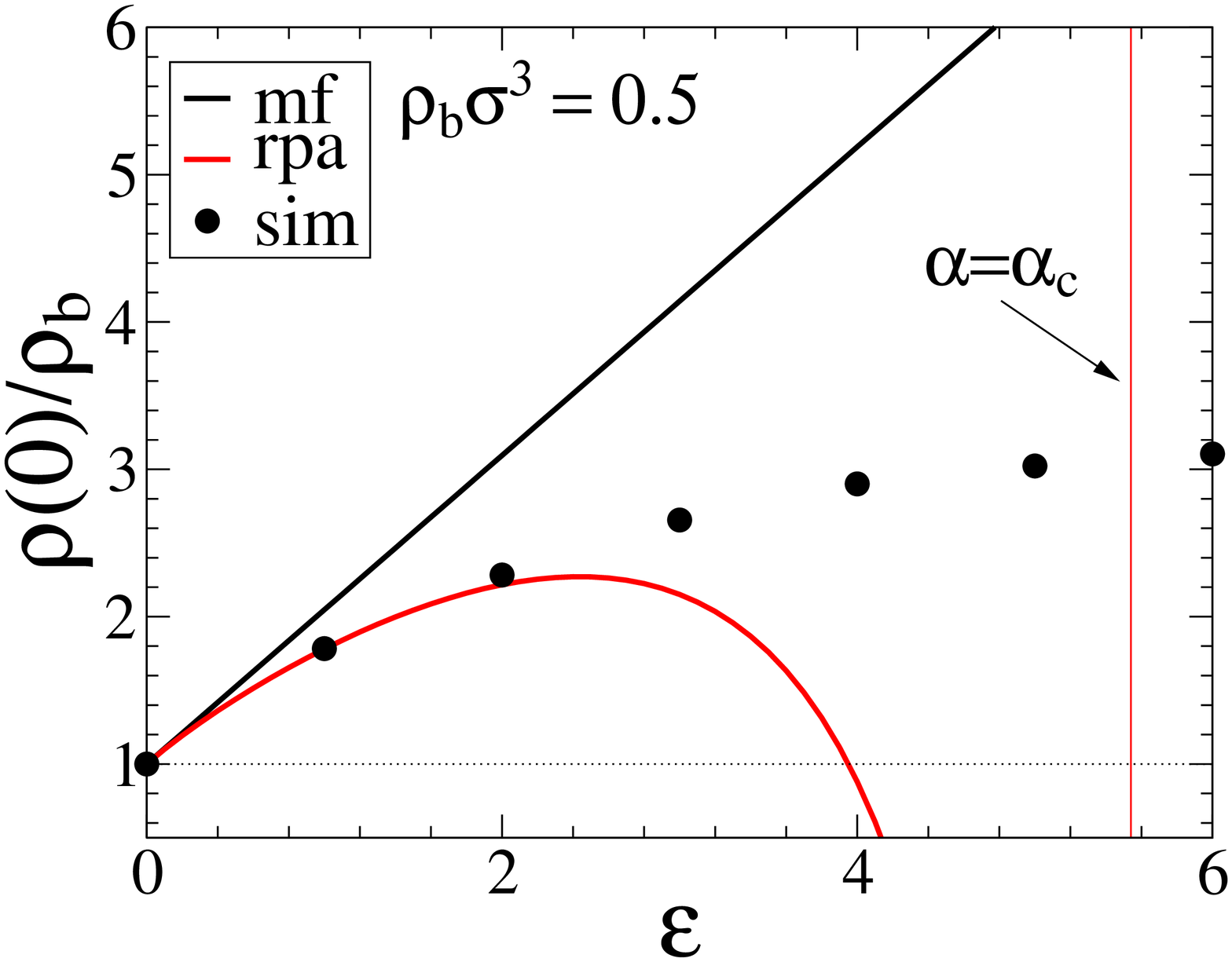}}
\caption{\label{fig:eta_1} 
Density at a contact with a planar wall, $\rho(0)$, as a function of an 
interaction strength $\varepsilon$ for penetrable-spheres.  The horizontal dotted line traces the ideal-gas behavior $\rho(0)/\rho_b=1$.    }
\end{figure}

For the final illustration of the performance of the RPA we calculate pressure with results shown in Fig. (\ref{fig:eta_1}).  
Because pressure is related to a density at a contact with a planar wall, $\beta P=\rho(0)$, the results provide insight about 
inhomogeneous density.  Within the RPA the pressure is given by 
\be
\frac{\beta P_{\rm}}{\rho_b} = 1+ \frac{\alpha}{6} + \frac{1}{2}\int_0^{1}d\lambda\,\frac{h_b^{\lambda}(0)-\lambda h_b(0)}{\lambda}, 
\label{eq:P1}
\ee
and is obtained from the free energy of a homogeneous system $P=-(\partial F/\partial V)_{T,N}$.  
The data points for pressure are generally more reliable than those for $\beta E$ in Fig. (\ref{fig:E5}).  The mean-field and the RPA, 
not designed to recover a correct dilute limit, should generally become more reliable for dense systems.  Going from $\rho_b\sigma^3=0.1$ 
to $\rho_b\sigma^3=0.3$ there is some improvement, but as
$\rho_b$ continues to increase, the reliability becomes limited by the presence of a singularity (or a solid like structure of correlations) 
at $\alpha_c\approx 34.81$.  The presence of the anomalous transition, therefore, limits the application of the RPA for penetrable-spheres.  
%%%%%%%%%%%%%%%%%%%%%%

\section{Planar wall confinment -- an inhomogeneous case}
\label{sec:inhom1}

\begin{figure}[h] 
 \begin{center}
 \begin{tabular}{rr}
  \includegraphics[height=0.2\textwidth,width=0.24\textwidth]{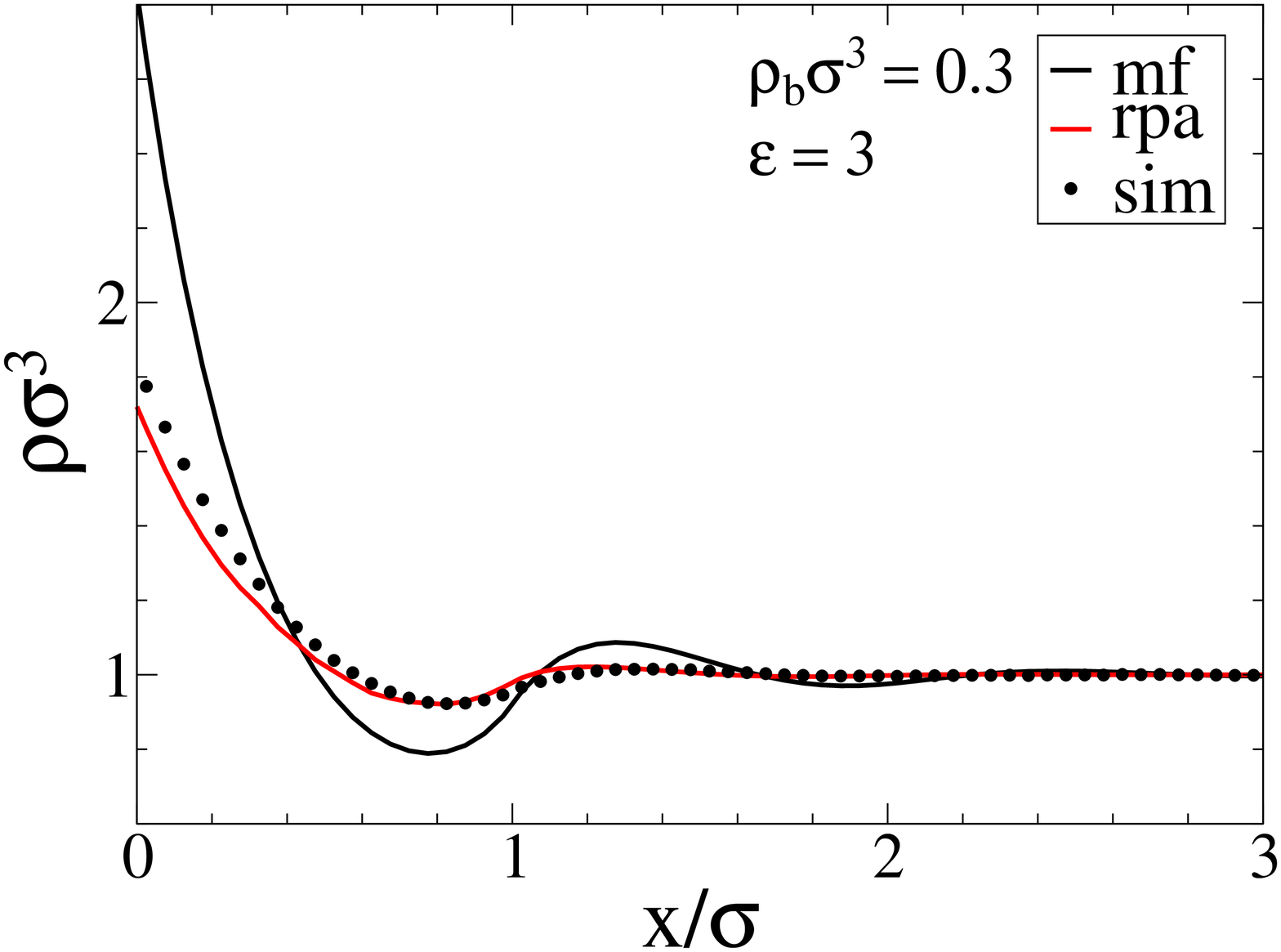}&
  \includegraphics[height=0.2\textwidth,width=0.24\textwidth]{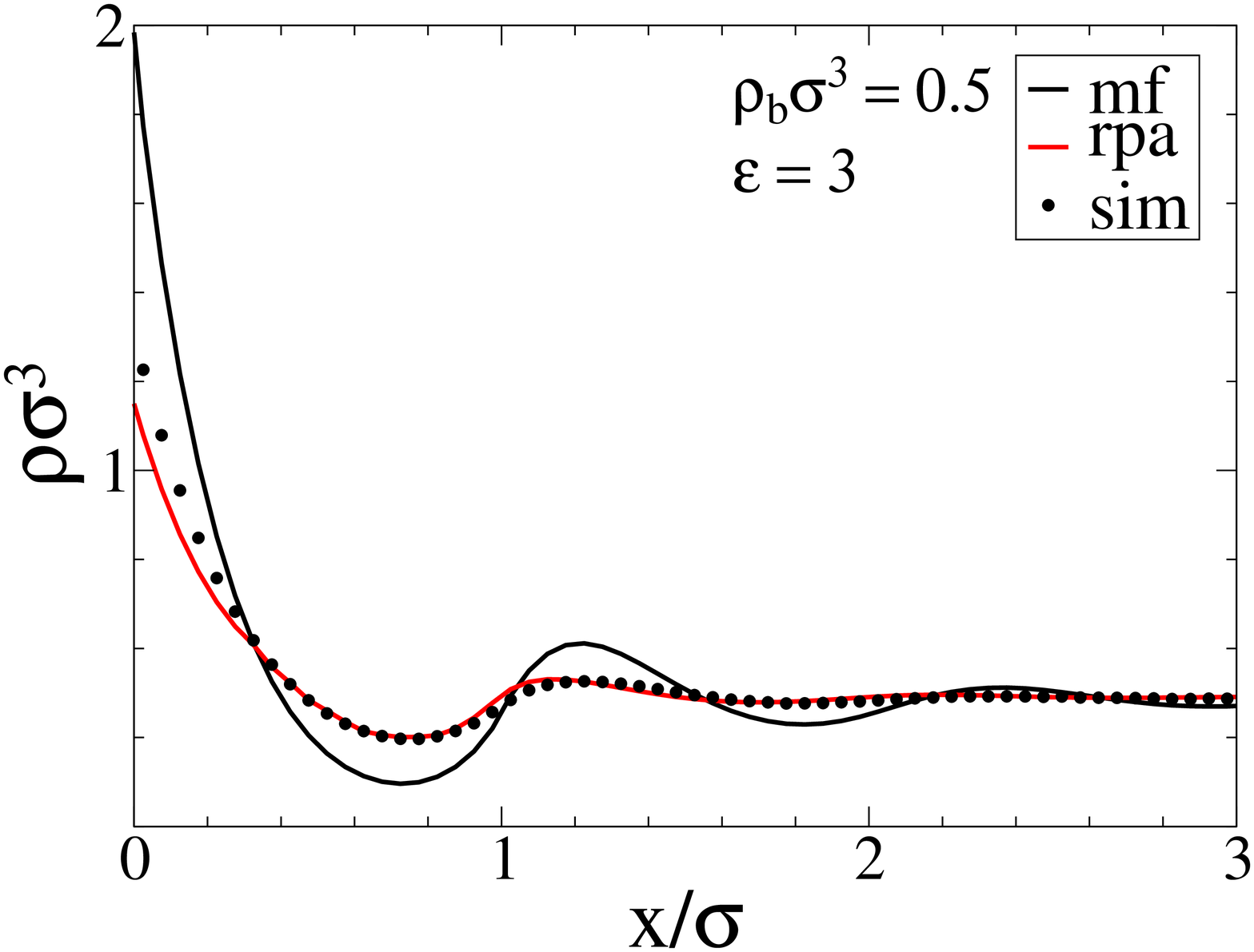}\\
 \end{tabular}
 \end{center}
\caption{Density profiles for penetrable-spheres confined by a wall at $x=0$.  } 
\label{fig:rho}
\end{figure}

In Fig. (\ref{fig:rho}) we plot the entire density profile for a fluid confined by a hard wall at $x=0$.  
The two profiles correspond to $\rho_b\sigma^3=0.3$ and $0.5$, for the interaction strength $\varepsilon=3$.  
For $\varepsilon>3$ the RPA starts to break down, see Fig. (\ref{fig:eta_1}).  
The RPA profile show decisive improvement in comparison to the bare mean-field.   
%%%%%%%%%%%%%%%%%%%%%
\graphicspath{{figures/}}
%%%%%%%%%%%%%%%%%%%%%

\begin{figure}[h] 
 \begin{center}
 \begin{tabular}{rr}
  \includegraphics[height=0.2\textwidth,width=0.26\textwidth]{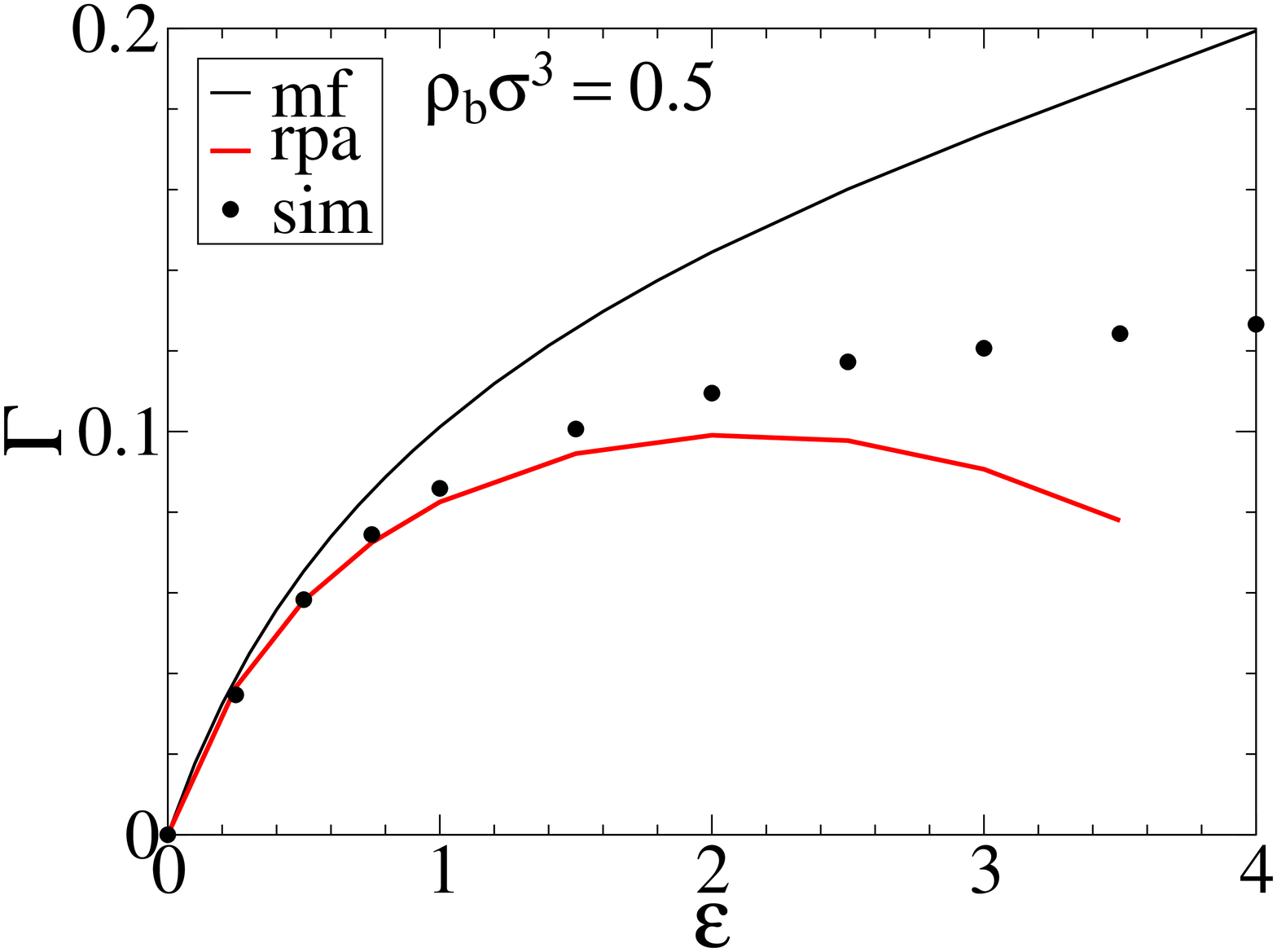}\\
 \end{tabular}
 \end{center}
\caption{Excess adsorption defined in Eq. (\ref{eq:gamma}) as a function of $\varepsilon$ for penetrable-spheres
near a hard-wall for the bulk density $\rho_b\sigma^3=0.5$.  } 
\label{fig:Gamma}
\end{figure}

Another quantity that offers a test of performance is the excess adsorption defined as
\be
\Gamma = \int_0^{\infty}dx\,\Big[\rho(x)-\rho_b\Big].  
\label{eq:gamma}
\ee
Unlike the contact density that is related to properties within a bulk, there's no equivalent relation available for the 
excess adsorption.  $\Gamma$ is the surface quantity and as such is related to a surface tension $\gamma$, or more 
specifically to the derivative of the surface tension w.r.t. the chemical potential $\mu$, 
\be
\Gamma = -\bigg(\frac{\partial\gamma}{\partial\mu}\bigg)_{T,V},
\ee
according to the Gibbs adsorption theorem.  In Fig. (\ref{fig:Gamma}) we plot $\Gamma$ for a single density $\rho_b\sigma^3=0.5$.  
As for the contact density, the RPA undermines $\Gamma$ while the mean-field systematically overestimates it.  
%%%%%%%%%%%%%%%%%%%%%
\graphicspath{{figures/}}
%%%%%%%%%%%%%%%%%%%%%

\section{Alternative approaches}
\label{sec:alter}

As the RPA applied to penetrable-spheres is unreliable both in the dilute limit (because of construction) and for large 
$\alpha$ (because of the emergence of a local solid-like structure for $\alpha\approx 34.81$), one may wish for 
an alternative approach.  To speculate such an approach, it may be helpful to consider a virial expansion of the excess 
free energy, 
\ba
F_{\rm ex} &=& \frac{1}{2}\int \!\!d{\bf r}_1\!\!\int\!\! d{\bf r}_2\,\rho({\bf r}_1)\rho({\bf r}_2)\bar f(r_{12})
\nonumber\\&+&
\frac{1}{6}\int \!\!d{\bf r}_1\!\!\int \!\!d{\bf r}_2\!\!\int \!\!d{\bf r}_3\,\rho({\bf r}_1)\rho({\bf r}_2)\rho({\bf r}_3)
\bar f(r_{12})\bar f(r_{23})\bar f(r_{31})
\nonumber\\&+&\dots, 
\label{eq:Fex}
\ea
where $\bar f(r)=1-e^{-\beta u_{}(r)}$ is the negative Mayer $f$-function, for penetrable-spheres given by 
$\bar f(r)=(1-e^{-\varepsilon})\Theta(\sigma-r)$.  The excess free energy incorporates all the contributions 
due to particle interactions, $F_{\rm ex} = \frac{1}{2}\int d{\bf r}\int d{\bf r}'\,\ra\rb\ua + F_c$. 
The two initial terms of the expansion have a simple ring topology.  The higher order terms, however, become
increasingly more complex, and the next term already involves three topologically distinct diagrams.

\begin{figure}[h] 
 \begin{center}
 \begin{tabular}{rr}
  \includegraphics[height=0.18\textwidth,width=0.2\textwidth]{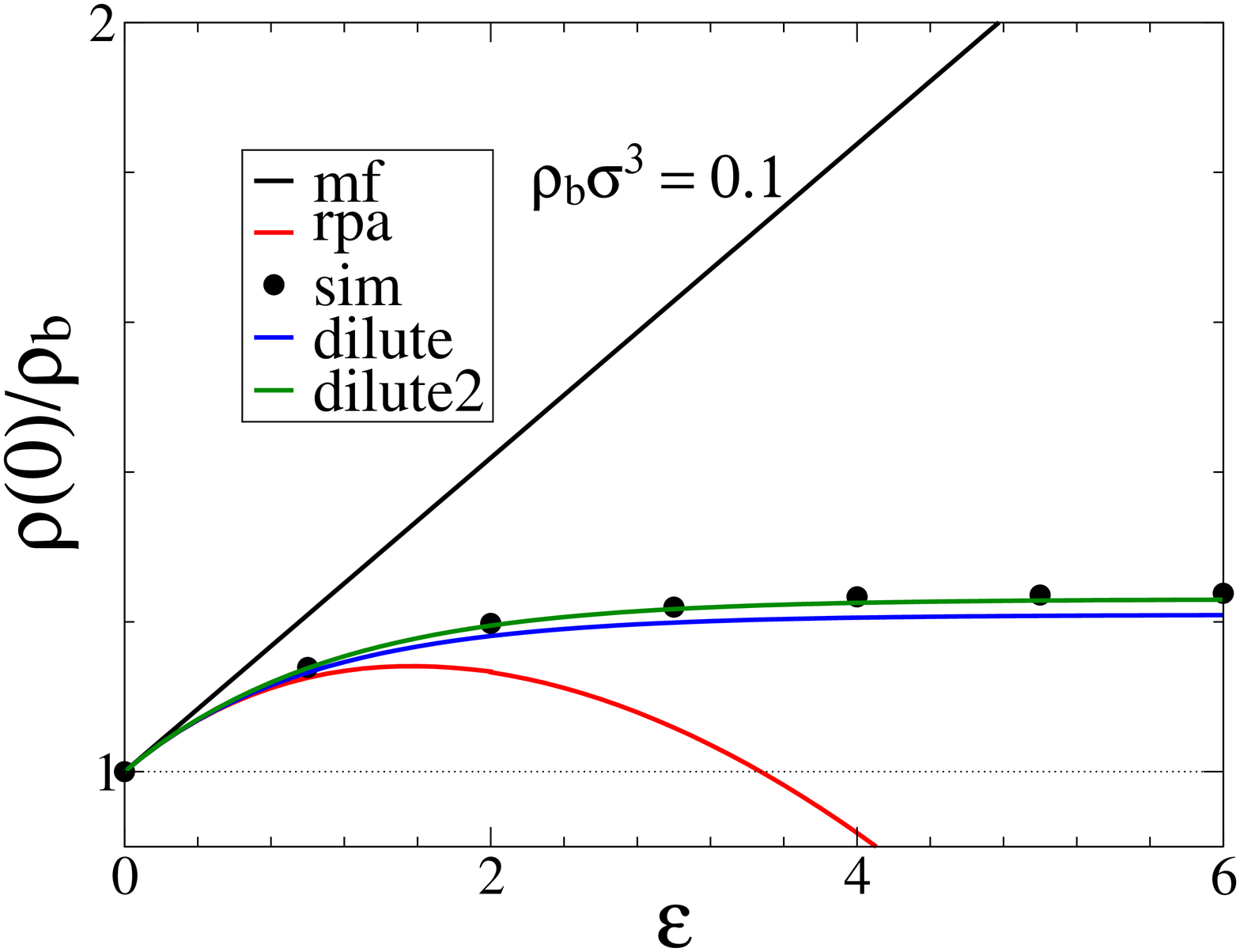}&
  \includegraphics[height=0.18\textwidth,width=0.2\textwidth]{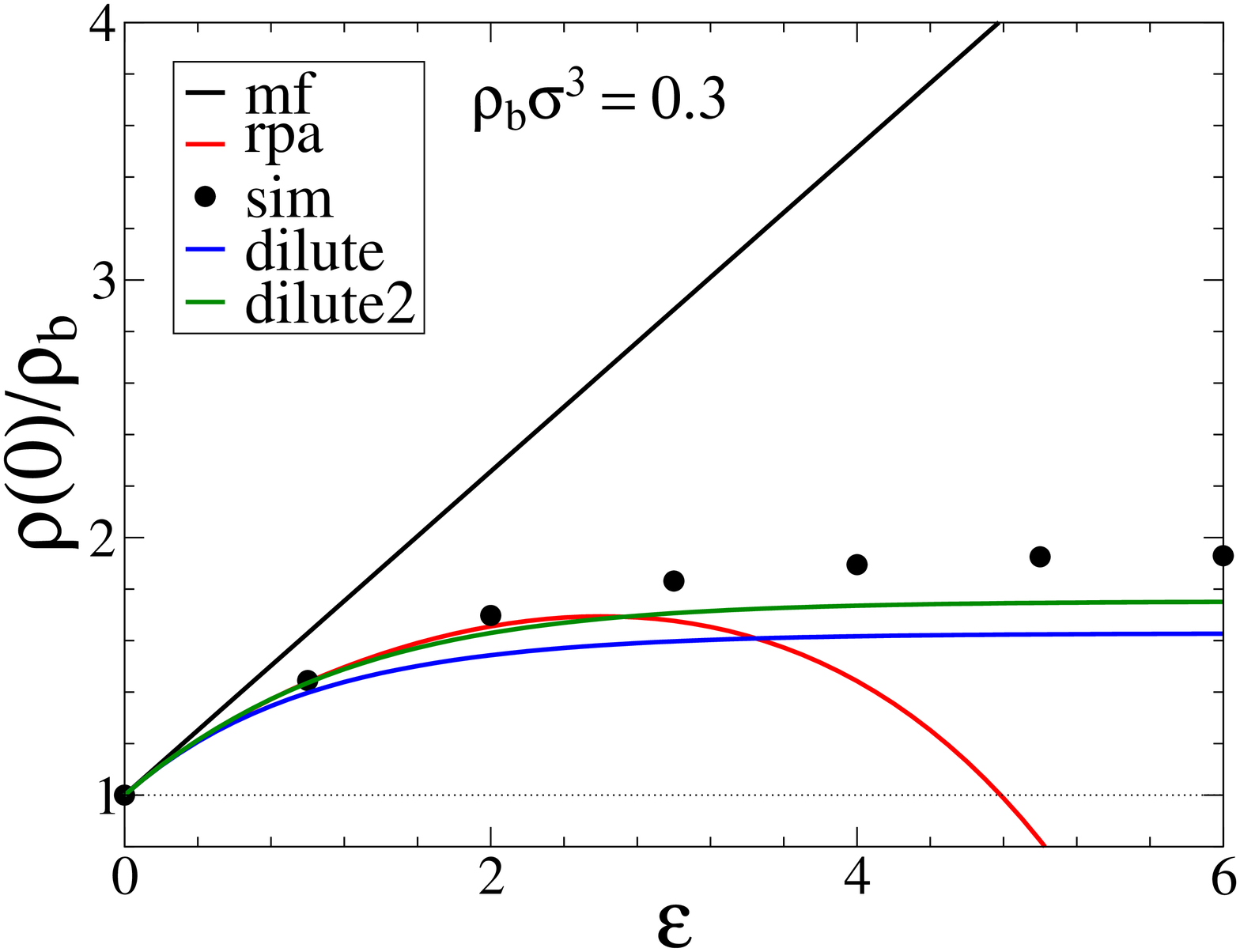}\\
 \end{tabular}
 \end{center}
\caption{Density at a contact with a planar wall, $\rho(0)$, as a function of an interaction strength $\varepsilon$ for penetrable-spheres. 
The results labeled as "dilute" correspond to Eq. (\ref{eq:P_dil}). } 
\label{fig:eta_c}
\end{figure}

In the dilute limit the first term in Eq. (\ref{eq:Fex}) should generally suffice,
\be
F_{\rm dil} = F_{\rm id} + \int d{\bf r}\,\rho({\bf r})U({\bf r}) 
+ \frac{1}{2}\int \!\!d{\bf r}_1\!\!\int\!\! d{\bf r}_2\,\rho({\bf r}_1)\rho({\bf r}_2)\bar f(r_{12}), 
\label{eq:Fdil}
\ee
and the resulting expression structurally resembles the mean-field approximation in Eq. (\ref{eq:Fmf}).  However, 
different parametrization of the two approaches corresponds to different limits.  Within the dilute limit approximation 
the pressure becomes 
\be
\frac{\beta P}{\rho_b} = 1 + \frac{4\pi\rho_b\sigma^3(1-e^{-\varepsilon})}{6}.
\label{eq:P_dil}
\ee
In Fig. (\ref{fig:eta_c}) we compare the above expression with the mean-field, the RPA, and the exact data points.  
The dilute limit approach systematically underestimates particle interactions and performs poorly for higher densities.  
%%%%%%%%%%%%%%%%%%%%%
\graphicspath{{figures/}}
%%%%%%%%%%%%%%%%%%%%%

Consecutive addition of terms of the virial expansion can correct the dilute limit predictions up to some point, but a truncated
series will always diverge at a sufficiently high density.  Instead of such a simple truncation, a "wiser" approach is to selectively 
include/exclude entire classes of diagrams.  Such a selection is naturally achieved by an application of more advanced closures, 
for example the Percus-Yevick or the hypernetted chain approximation.  However, when combined with the adiabatic connection 
framework, Eq. (\ref{eq:Fc}) and Eq. (\ref{eq:OZ_lambda}), these closures introduce another level of complexity as 
the functional derivative $\frac{\delta F_{\rm ex}}{\delta\rho}$ involves an integral over $\lambda$, and to calculate a  
density, correlations for all values of $\lambda$, from $0$ to $1$, are needed.  This difficulty precludes
these approaches from having practical use.

To take a different approach, we take advantage of the RPA mathematical framework which is, as said before, defined 
in terms of integrals with ring topology.  Defining the operator $B({\bf r},{\bf r}')=\rho({\bf r})\bar f({\bf r},{\bf r}')$ and using 
conventions of the linear algebra, all the ring terms of the excess free energy can be represented as 
\ba
F_{\rm ex}^{\rm dil2}[\rho] &=& \frac{1}{2}\int d{\bf r}\int d{\bf r}'\,\rho({\bf r})\rho({\bf r}')\bigg[\bar f({\bf r},{\bf r}')
+\frac{\bar f^2({\bf r},{\bf r}')}{2}\bigg]\nonumber\\ 
&+&\frac{1}{2}{\rm Tr}\bigg[\log(I+B) - B\bigg].
\ea
We compare this with the RPA excess free energy given by
\ba
F_{\rm ex}^{\rm rpa}[\rho] &=& \frac{1}{2}\int d{\bf r}\int d{\bf r}'\,\rho({\bf r})\rho({\bf r}')u({\bf r},{\bf r}')
\nonumber\\ &+&\frac{1}{2}{\rm Tr}\bigg[\log(I+A) - A\bigg],
\ea
where $A({\bf r},{\bf r}')=\rho({\bf r})u({\bf r},{\bf r}')$.  There are two main differences between the two expressions.  
First, $\ua$ is substituted by $\bar f({\bf r},{\bf r}')$.  Second, the "mean-field" part involves an extra term.  
In Fig. (\ref{fig:eta_c}) we plot the pressure data points from this new scheme labeled as "dilute2".  
There is a slight improvement over the results in Eq. (\ref{eq:P_dil}), but taking into account the additional complexity, 
the improvement does not justify the approach. It becomes clear that terms with more complex topology are necessary 
for an accurate description of dense fluids.

\section{Two-component system}
\label{sec:hom2}

In this section we consider a two-component system with symmetric interactions 
defined as follows:  particles of the same species interact as $u(r<\sigma)=\varepsilon$, 
and particles of different species interact as $u(r<\sigma)=-\varepsilon$, in both cases $u(r>\sigma)=0$.   
%\[ u_{ij}(r<\sigma) = \left\{ 
%  \begin{array}{r l}
%     \varepsilon , & \quad \text{if $i=j$}\\
%    -\varepsilon, & \quad \text{if $i\ne j$.}
%\label{eq:u}
%  \end{array} \right.\]
%where 
This is an interesting scenario because all the mean-field contributions cancel out and interactions are captured 
only through correlational contributions.   

The present system gives rise to four different correlations, $h_{11}({\bf r},{\bf r}'),h_{22}({\bf r},{\bf r}'),h_{12}({\bf r},{\bf r}'),h_{21}({\bf r},{\bf r}')$, 
where $1$ and $2$ designate two  species.  By virtue of the symmetry of the interactions $h_{11}({\bf r},{\bf r}')=h_{22}({\bf r},{\bf r}')$
and $h_{12}({\bf r},{\bf r}')=h_{21}({\bf r},{\bf r}')$.  Furthermore, within the RPA $h_{11}({\bf r},{\bf r}')=-h_{21}({\bf r},{\bf r}')$, 
$h_{22}({\bf r},{\bf r}')=-h_{21}({\bf r},{\bf r}')$, and so forth, so that all correlations may be represented by a single 
functional form $h^{\rm }({\bf r},{\bf r}')$ obtained from the OZ relation 
\be
h({\bf r },{\bf r}') = -\beta u({\bf r},{\bf r}') - \beta\!\!\int \!\!d{\bf r}''\rho({\bf r}'')h({\bf r}',{\bf r}'')u({\bf r},{\bf r}''), 
\ee
where $\ra = \rho_1({\bf r}) + \rho_2({\bf r})$ is the total number density, 
$h_{11}({\bf r},{\bf r}')=h({\bf r},{\bf r}')$, $h_{12}({\bf r},{\bf r}')=-h({\bf r},{\bf r}')$, 
and so forth.  Now, if such a two-component fluid is constrained to a half-space $x>0$ by a planar wall, a number 
density for a species $i$ is given by 
\be
\rho_i({\bf r}) = \frac{\rho_{b}}{2} e^{\frac{1}{2}[h({\bf r},{\bf r})-h_b(0)]},
\label{eq:rho_i}
\ee
where $\rho_b$ is the total bulk density.  Note that the mean-field contributions are absent and the density 
is determined solely by correlations.

\begin{figure}[h] 
 \begin{center}
 \begin{tabular}{rr}
  \includegraphics[height=0.2\textwidth,width=0.28\textwidth]{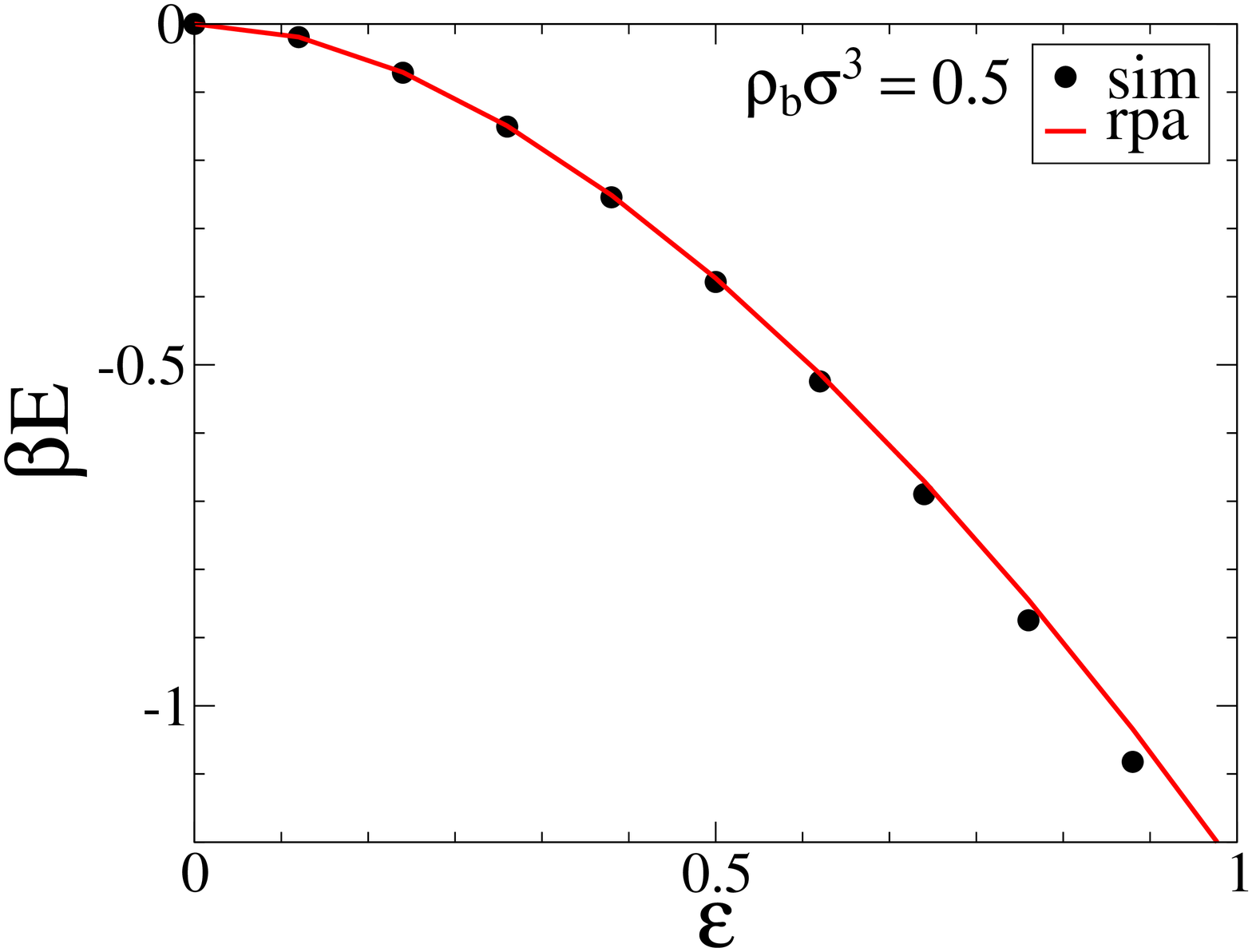}\\
 \end{tabular}
 \end{center}
\caption{Average potential energy per particle as a function of $\varepsilon$ for a homogenous two-component system with total density 
$\rho_b\sigma^3=0.5$.  The RPA result corresponds to Eq. (\ref{eq:E_rpa2}).  There are no mean-field contributions.} 
\label{fig:E52}
\end{figure}

\begin{figure}[h] 
 \begin{center}
 \begin{tabular}{rrc}
  \includegraphics[height=0.16\textwidth,width=0.2\textwidth]{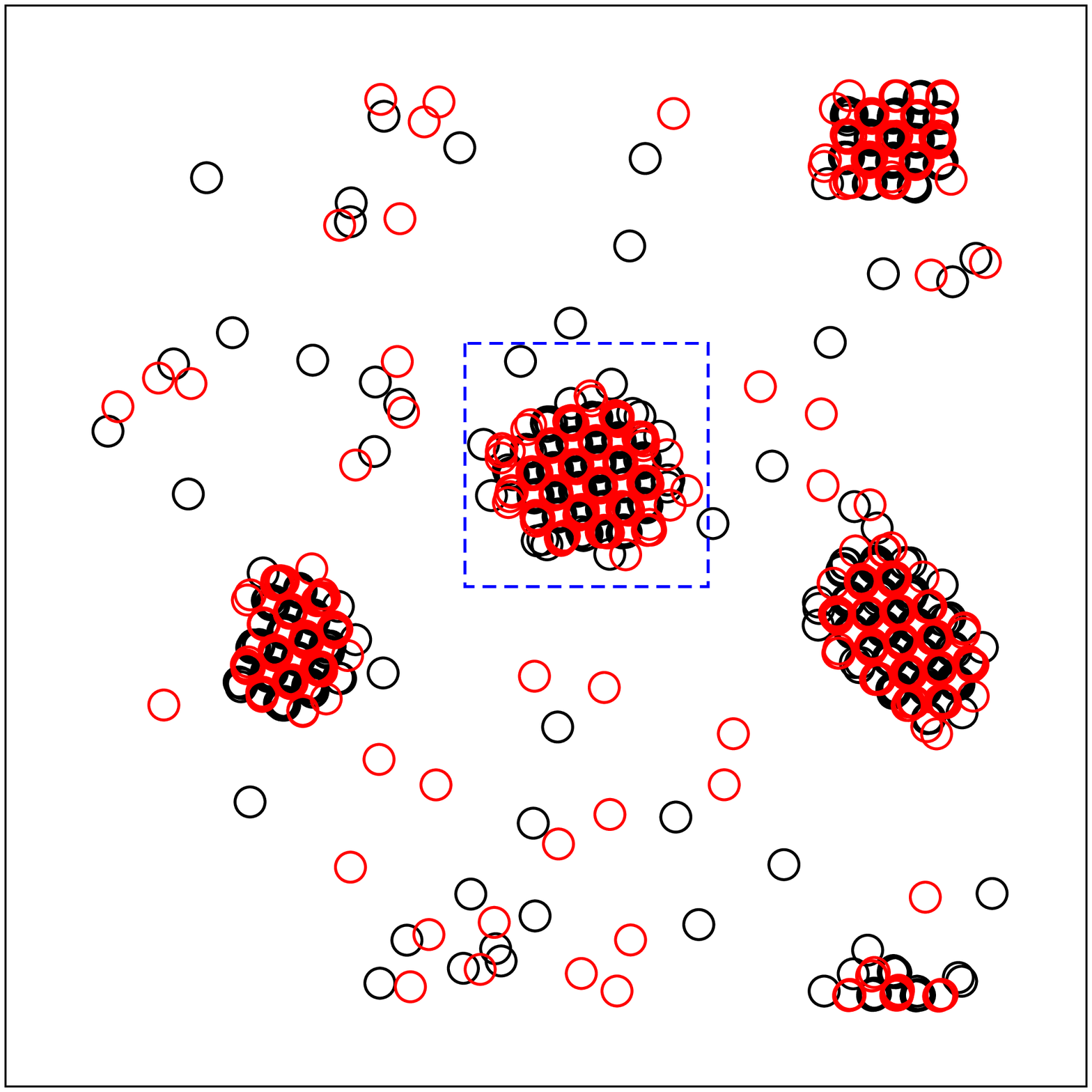}&
% \hspace{0.5cm}
  \includegraphics[height=0.16\textwidth,width=0.2\textwidth]{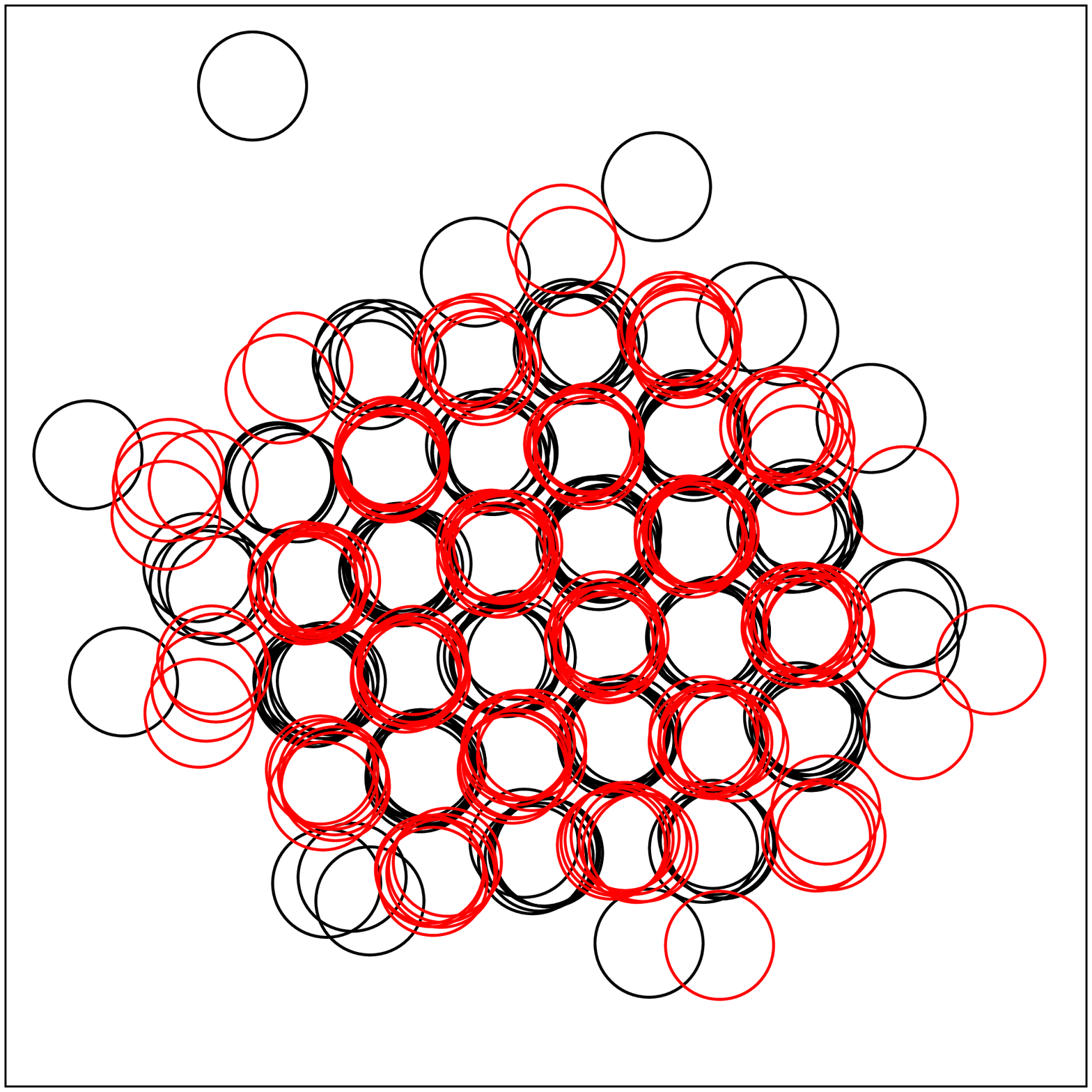}\\
 \end{tabular}
 \end{center}
\caption{Configurational snapshots of a two-component penetrable-sphere system, after the phase transition. For the sake of 
simplicity the snapshots are for a 2D system. The red and black circles indicate particles of different species. The system 
parameters are $\varepsilon=1$ and $\rho_b\sigma^2=0.9$.  The first figure contains $N=1000$ particles and the second one $N=240$.  } 
\label{fig:solid}
\end{figure}

Considering a homogeneous fluid, an average potential energy that a particle of the species $1$ feel is
\be
\beta E  = \frac{\varepsilon\rho_b}{2}\int_0^{\sigma} dr\,r^2 \big[h_{11}(r)-h_{12}(r)\big].
\label{eq:E_2}
\ee
Within the RPA the above expression can be written as  
\be
\beta E^{\rm rpa} = -h_b^{\rm rpa}(0)-\varepsilon.  
\label{eq:E_rpa2}
\ee
Fig. (\ref{fig:E52}) shows $\beta E$ as a function of $\varepsilon$ for $\rho_b\sigma^3=0.5$.  Within the range 
$\varepsilon<1$ the RPA data points are highly accurate.  
%%%%%%%%%%%%%%%%%%%%%
\graphicspath{{figures/}}
%%%%%%%%%%%%%%%%%%%%%
The fact that $\beta E$ monotonically decreases with increasing $\varepsilon$ indicates increased 
association between particles of different species, which eventually leads to phase-transition around $\varepsilon=1$.  
For illustration in Fig. (\ref{fig:solid}) we show a configuration snapshot of a two-component system in 2D after
phase transition where the system consists of highly structured clusters.   The RPA does not indicate any phase transition
around this point.  
%%%%%%%%%%%%%%%%%%%%%
\graphicspath{{figures/}}
%%%%%%%%%%%%%%%%%%%%%

\begin{figure}[h] 
 \begin{center}
 \begin{tabular}{rrc}
  \includegraphics[height=0.18\textwidth,width=0.22\textwidth]{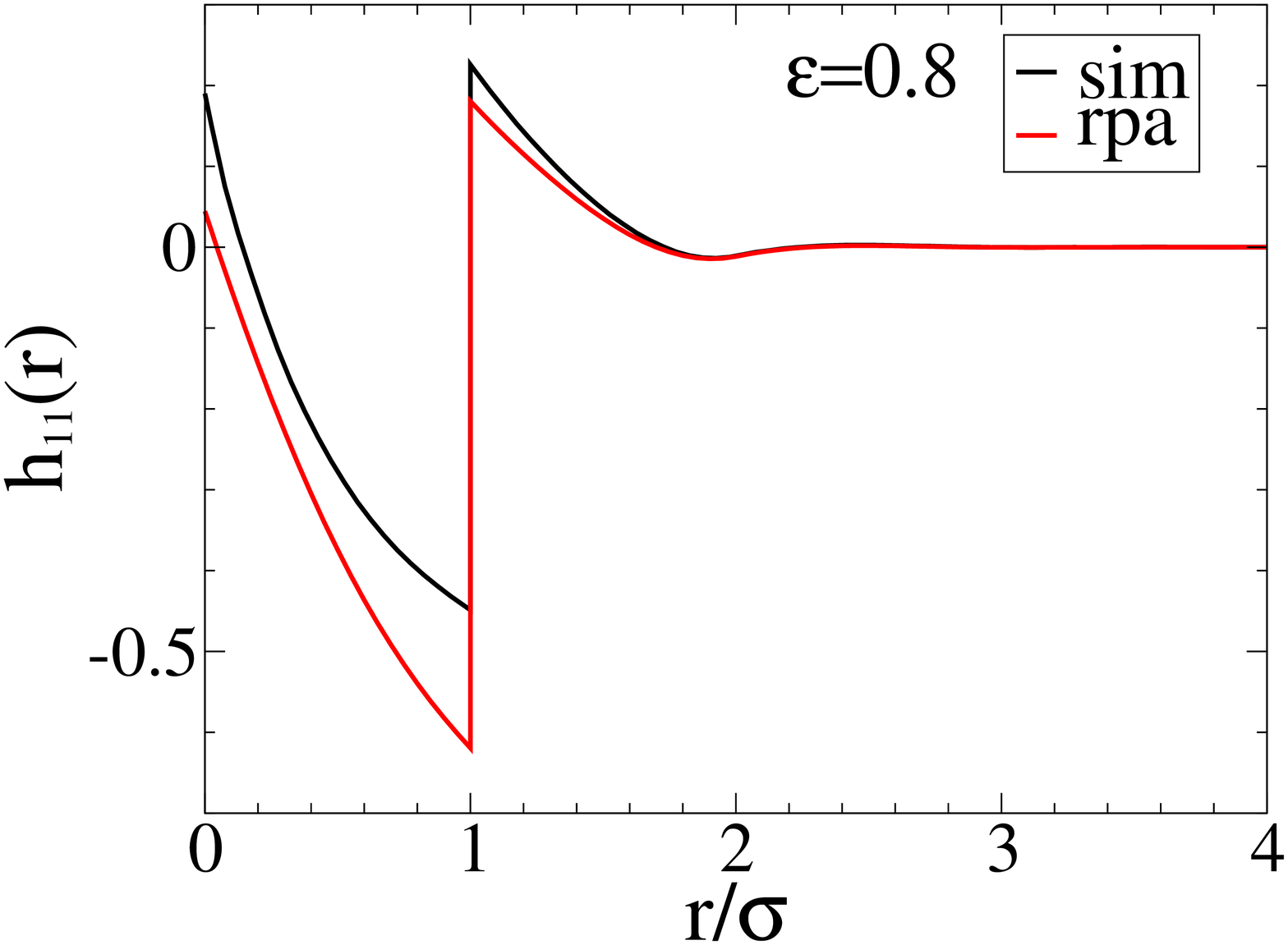}&
  \includegraphics[height=0.18\textwidth,width=0.22\textwidth]{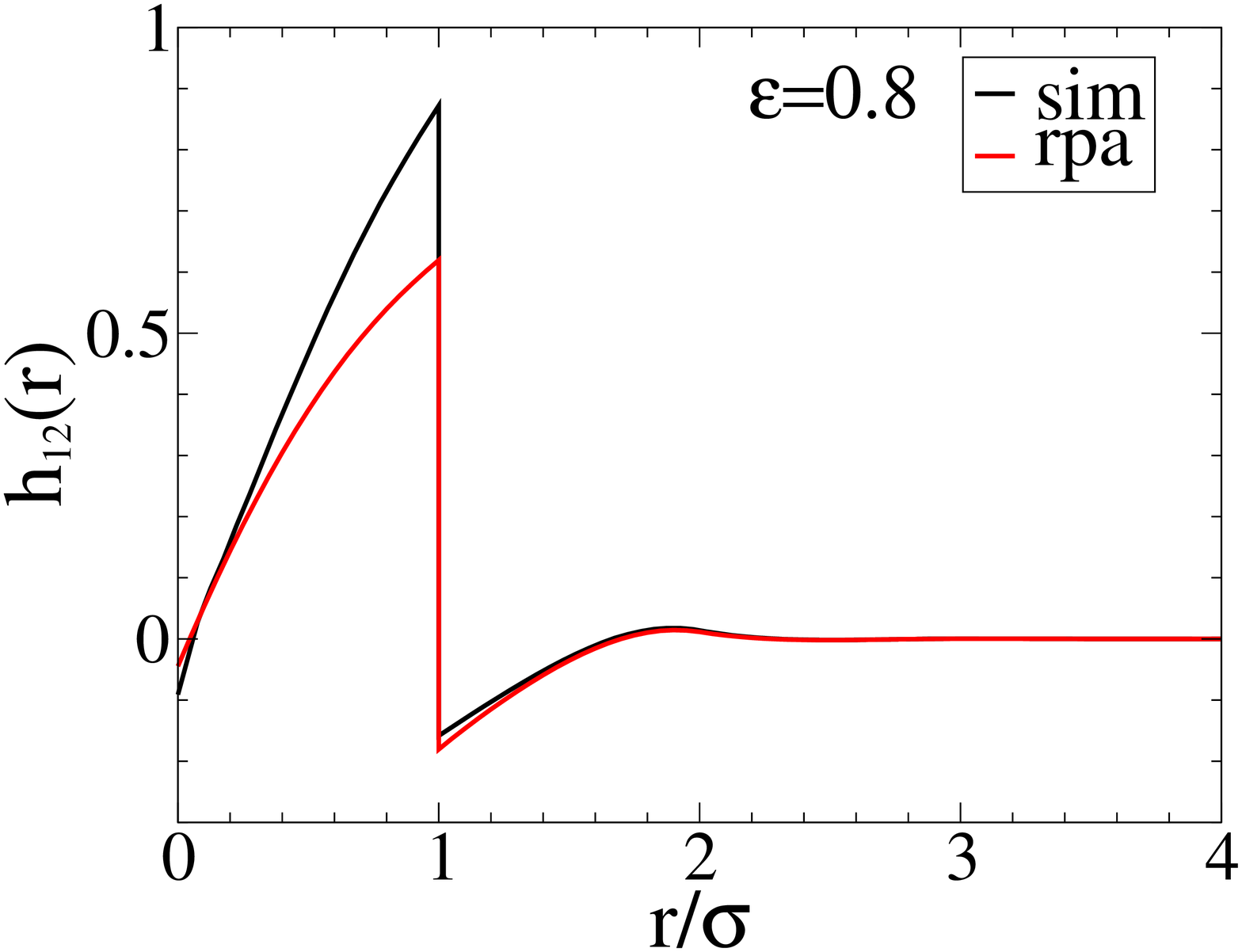}\\
 \end{tabular}
 \end{center}
\caption{Pair correlation functions of a homogeneous two-component fluid for $\rho_b\sigma^3=0.5$.  By virtue of symmetric interactions,
$h_{11}({\bf r},{\bf r}')=h_{22}({\bf r},{\bf r}')$ and $h_{12}({\bf r},{\bf r}')=h_{21}({\bf r},{\bf r}')$. } 
\label{fig:hr08a}
\end{figure}

\begin{figure}[h] 
 \begin{center}
 \begin{tabular}{rr}
  \includegraphics[height=0.18\textwidth,width=0.22\textwidth]{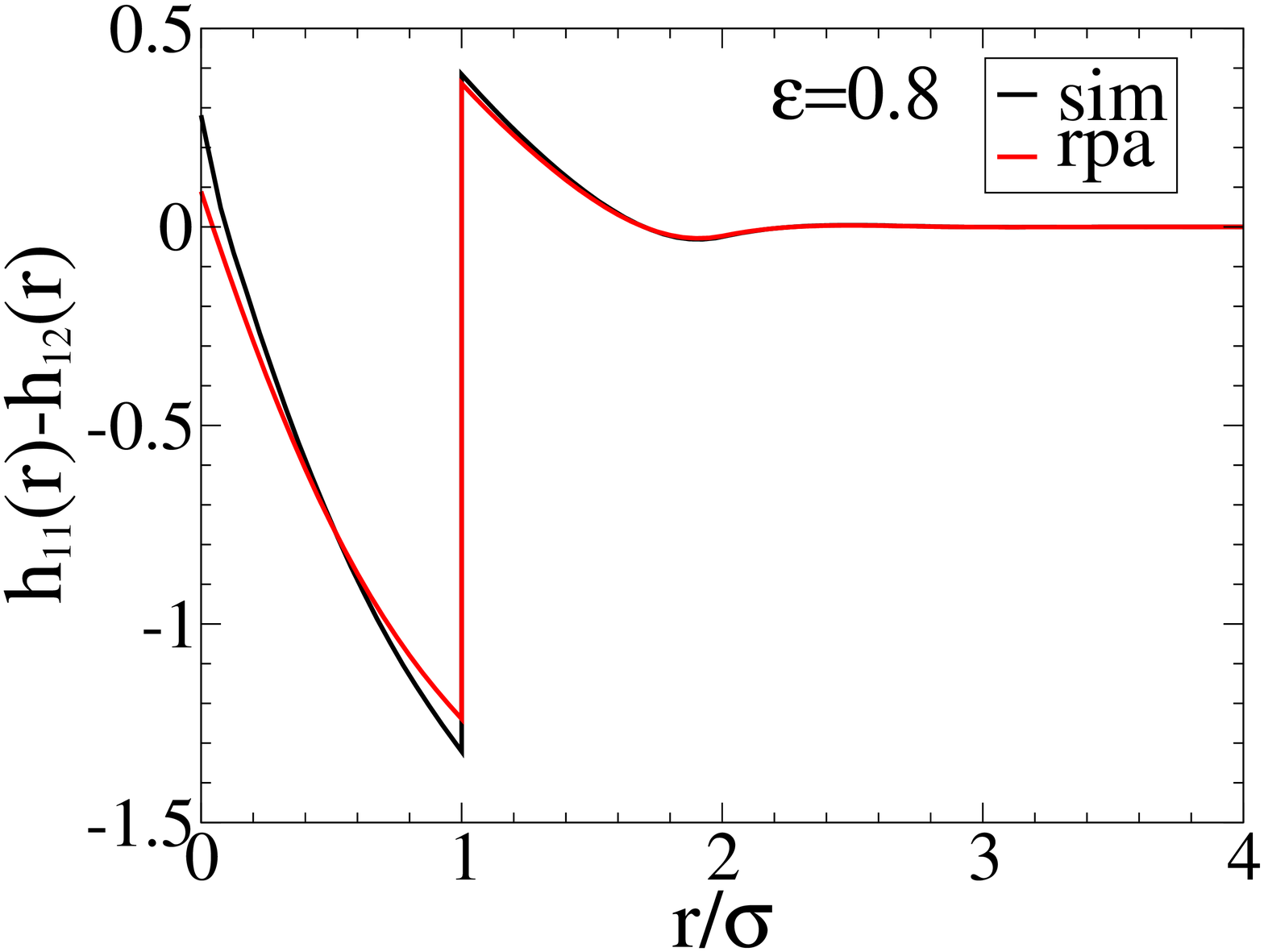}&
 \end{tabular}
 \end{center}
\caption{Correlations $h_{11}(r)-h_{12}(r)$ of a homogeneous fluid for $\rho_b\sigma^3=0.5$.  Compare with Fig. (\ref{fig:hr08a}) to 
understand the cancellation of errors in $h_{11}(r)$ and $h_{12}(r)$. } 
\label{fig:hr08b}
\end{figure}

%To understand the performance of the RPA, 
In Fig. (\ref{fig:hr08a}) we look into correlations $h_{11}(r)$ and $h_{12}(r)$, 
which reveal that the RPA correlations are not at all that accurate.  
%%%%%%%%%%%%%%%%%%%%%
\graphicspath{{figures/}}
%%%%%%%%%%%%%%%%%%%%%
From Eq. (\ref{eq:E_2}) we know, however, that $\beta E$ depends on the quantity $h_{11}(r)-h_{12}(r)$ (within the RPA given by $2h(r)$).  
Accordingly, in Fig. (\ref{fig:hr08b}) we plot $h_{11}(r)-h_{12}(r)$.  Here the agreement between the simulation and the RPA is quite
impressive and indicates some cancellation of errors, accrued in both $h_{11}(r)$ and $h_{12}(r)$, that is responsible for good results
seen in Fig. (\ref{fig:E52}).  
%%%%%%%%%%%%%%%%%%%%%
\graphicspath{{figures/}}
%%%%%%%%%%%%%%%%%%%%%

\begin{figure}[h] 
 \begin{center}
 \begin{tabular}{rr}
  \includegraphics[height=0.2\textwidth,width=0.28\textwidth]{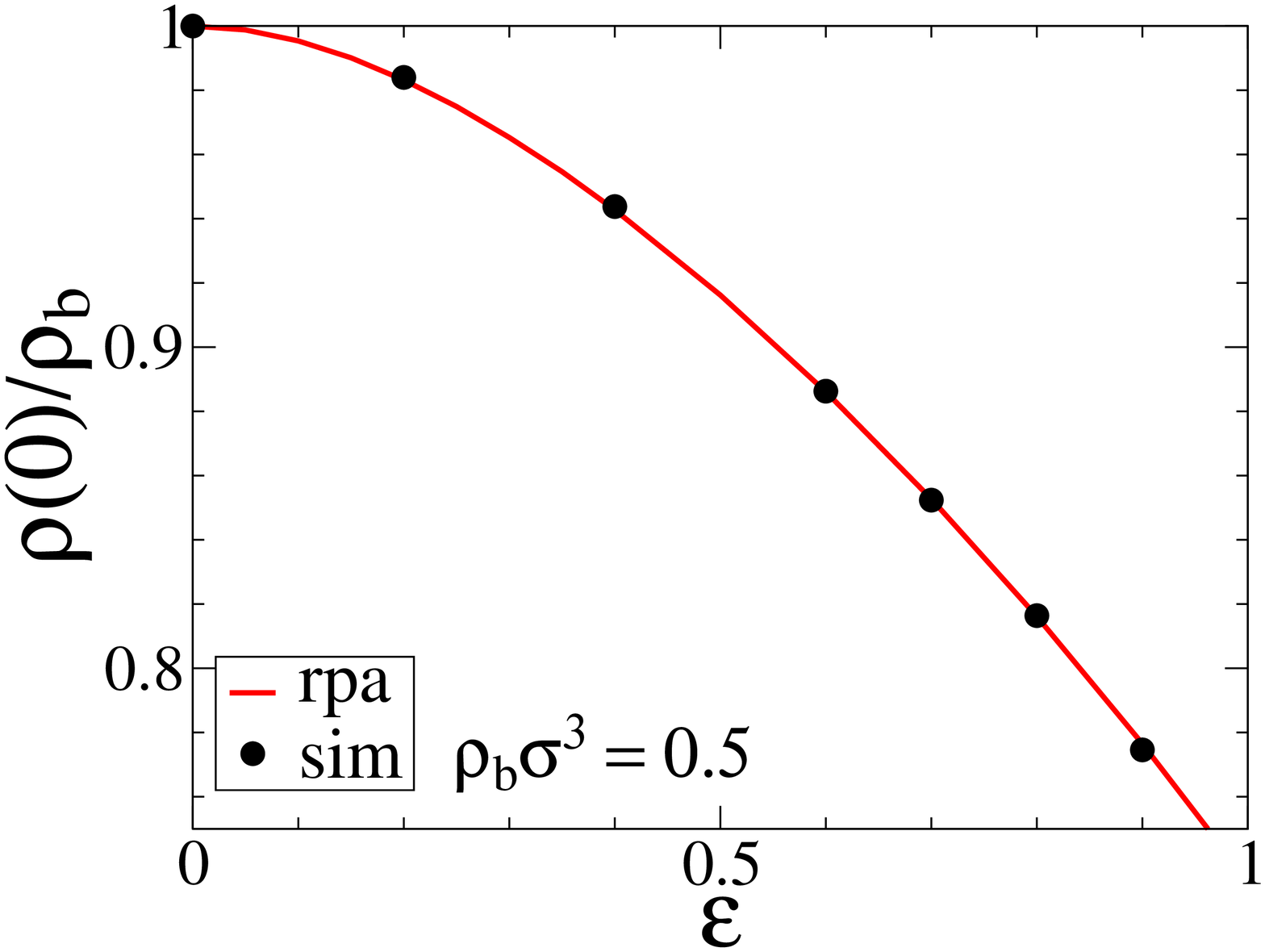}\\
 \end{tabular}
 \end{center}
\caption{Density at a contact with a planar wall, $\rho(0)=\rho_1(0)+\rho_2(0)$, as a function of an interaction strength $\varepsilon$. } 
\label{fig:eta2}
\end{figure}

\begin{figure}[h] 
 \begin{center}
 \begin{tabular}{rr}
  \includegraphics[height=0.2\textwidth,width=0.28\textwidth]{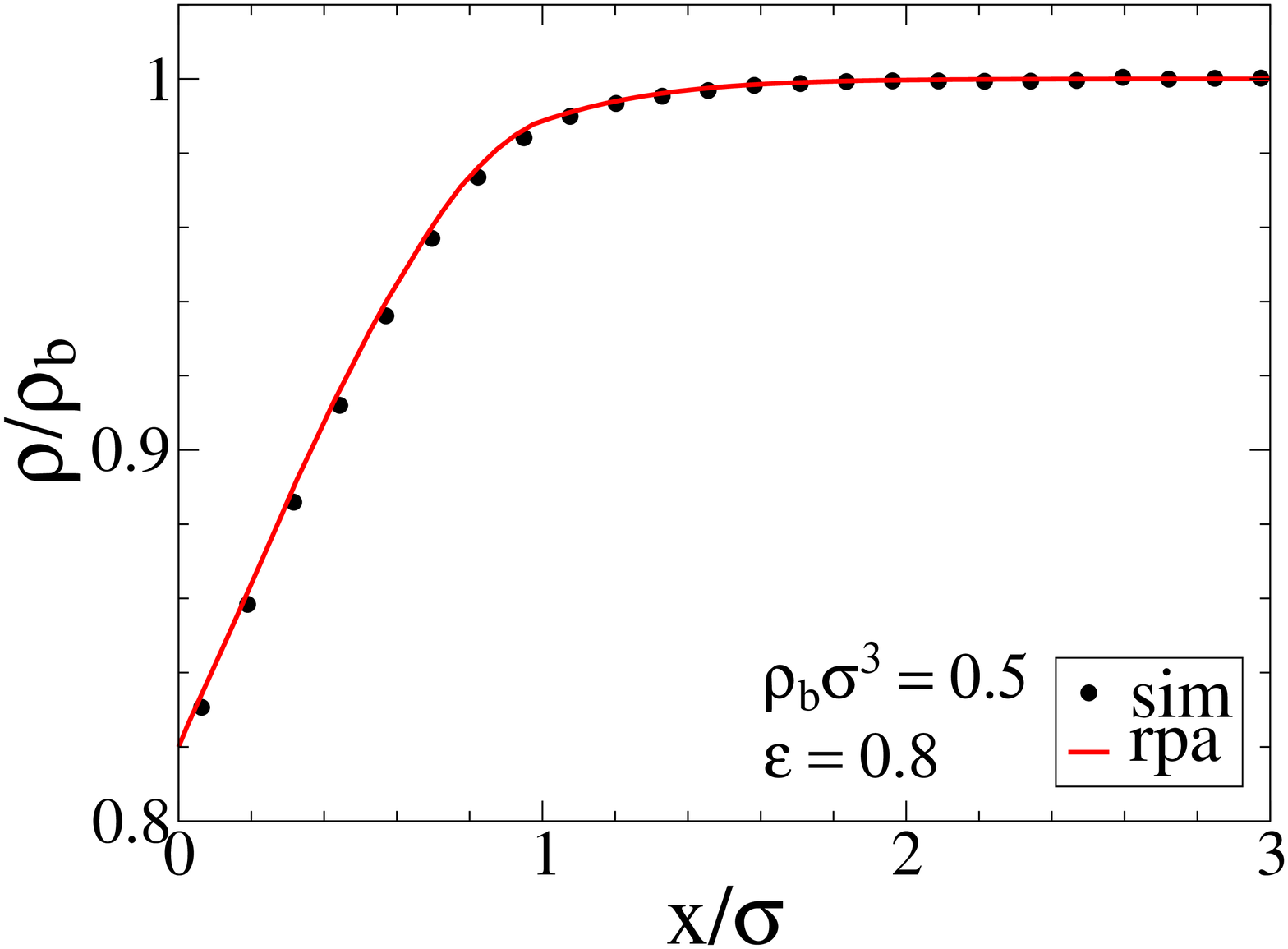}\\
 \end{tabular}
 \end{center}
\caption{Density profile for two-component penetrable-spheres near a planar wall at $x=0$. }
\label{fig:rho2}
\end{figure}

Next we calculate the pressure, within the RPA given by 
\be
\frac{\beta P_{\rm}}{\rho_b} = 1+ \frac{1}{2}\int_0^{1}d\lambda\,\bigg(\frac{h_b^{\lambda}(0)-\lambda h_b(0)}{\lambda}\bigg).
\ee
where 
\be
h_{\lambda}^{\rm rpa}({\bf r },{\bf r}') = - \lambda\beta u({\bf r},{\bf r}') -  \lambda\beta \!\!\int \!\!d{\bf r}''\rho({\bf r}'')h^{\rm rpa}_{\lambda}({\bf r}',{\bf r}'')u({\bf r},{\bf r}''), 
%\label{eq:OZ_RPA4}
\ee
The expression for pressure is similar to that in Eq. (\ref{eq:P1}) without the mean-field term.  The results are plotted in 
Fig. (\ref{fig:eta2}).  Once again, the agreement between the simulation and the RPA is quite excellent.  
%%%%%%%%%%%%%%%%%%%%%
\graphicspath{{figures/}}
%%%%%%%%%%%%%%%%%%%%%
Finally, in Fig. (\ref{fig:rho2}) we plot the entire density profiles for penetrable-spheres confined by a wall at $x=0$.  The RPA
for this situation once again turns out being very accurate.  
%%%%%%%%%%%%%%%%%%%%%
\graphicspath{{figures/}}
%%%%%%%%%%%%%%%%%%%%%

\section{Conclusion}
\label{sec:concl}

The main lesson to be taken from this study is that the RPA, despite its non-perturbative construction, is not a 
theory of the strong-coupling regime. The RPA introduces correlations to the mean-field level of description, but 
does not really extend the range of validity of the mean-field. The RPA is shown to be especially useful for situations 
in which the mean-field contributions are cancelled, as in the case of a two-component system considered in the 
present study. The RPA for this situation is found to be rather accurate, albeit within the weak-coupling limit, while 
the cause of this success lies in a somewhat lucky cancellation of errors.
%The main lesson to be taken from this study is that the RPA, despite its non-perturbative construction, is not a 
%theory of the strong-coupling regime.  The RPA introduces correlations to the mean-field level of description but 
%does not really extend the range of validity of the mean-field.  The RPA is shown to be especially useful for 
%situations in which the mean-field contributions are cancelled, as for the case of a two-component system 
%considered in the present study.  In this situation particle interactions can only be captured by correlations, as
%the mean-field term gets cancelled.  The RPA for this situation proves itself to be rather accurate, albeit within the 
%weak-coupling limit, and the cause of this success lies in a somewhat lucky cancellation of errors.  

%------------------------------------------------
\begin{acknowledgments}
D.F. acknowledges financial support by PNDP-Capes, under the project PNPD20132533.  
\end{acknowledgments}

%------------------------------------------------
% References
%------------------------------------------------

%------------------------------------------------

\end{document}